\documentclass[iop]{emulateapj}
\slugcomment{Published in the Astrophysical Journal, 848:37}
\usepackage[colorlinks,urlcolor=blue,citecolor=blue,linkcolor=blue,breaklinks=true]{hyperref} 
\usepackage{graphicx}
\usepackage{amsmath}

\usepackage{breakcites}

\shorttitle{CLASH Photometry}
\shortauthors{Connor et al.}
\begin{document}

\title{Crowded Field Galaxy Photometry: Precision Colors in the CLASH Clusters}
\author{Thomas Connor\altaffilmark{1,2,3}, Megan Donahue\altaffilmark{3}, Daniel D. Kelson\altaffilmark{1}, John Moustakas\altaffilmark{4}, Dan Coe\altaffilmark{5}, Marc Postman\altaffilmark{5}, Larry D. Bradley\altaffilmark{5}, Anton M. Koekemoer\altaffilmark{5}, Peter Melchior\altaffilmark{6},  Keiichi Umetsu\altaffilmark{7}, G. Mark Voit\altaffilmark{3}}

\altaffiltext{1}{The Observatories of the Carnegie Institution for Science, 813 Santa Barbara Street, Pasadena, CA 91101, USA}
\altaffiltext{2}{tconnor@carnegiescience.edu}
\altaffiltext{3}{Department of Physics and Astronomy, Michigan State University, East Lansing, MI 48823, USA}
\altaffiltext{4}{Department of Physics and Astronomy, Siena College, 515 Loudon Road, Loudonville, NY 12211, USA}
\altaffiltext{5}{Space Telescope Science Institute, 3700 San Martin Drive, Baltimore, MD 21208, USA}
\altaffiltext{6}{Department of Astrophysical Sciences, Princeton University, Peyton Hall, Princeton, NJ 08544, USA}
\altaffiltext{7}{Institute of Astronomy and Astrophysics, Academia Sinica, P.O. Box 23-141, Taipei 10617, Taiwan}
\begin{abstract}

We present a new method for photometering objects in galaxy clusters. We introduce a mode-filtering technique for removing spatially variable backgrounds, improving both detection and photometric accuracy (roughly halving the scatter in the red sequence compared to previous catalogs of the same clusters). This method is based on robustly determining the distribution of background pixel values and should provide comparable improvement in photometric analysis of any crowded fields. We produce new multiwavelength catalogs for the 25 CLASH cluster fields in all 16 bandpasses from the UV through the near IR, as well as rest-frame magnitudes. A comparison with spectroscopic values from the literature finds a ${\sim} 30\%$ decrease in the redshift deviation from previously released CLASH photometry. This improvement in redshift precision, in combination with a detection scheme designed to maximize purity, yields a substantial upgrade in cluster member identification over the previous CLASH galaxy catalog. We construct luminosity functions for each cluster, reliably reaching depths of \textit{at least} 4.5 mag below $M^*$ in every case, and deeper still in several clusters. We measure $M^*$, $\alpha$, and their redshift evolution, assuming the cluster populations are coeval, and find little to no evolution of $\alpha$, $-0.9\lesssim\langle\alpha\rangle\lesssim -0.8$, and $M^*$ values consistent with passive evolution. We present a catalog of galaxy photometry, photometric and spectroscopic redshifts, and rest-frame photometry for the full fields of view of all 25 CLASH clusters. Not only will our new photometric catalogs enable new studies of the properties of CLASH clusters, but mode-filtering techniques, such as those presented here, should greatly enhance the data quality of future photometric surveys of crowded fields.

\end{abstract}

\section{Introduction}

Current models of structure formation predict that rich clusters of galaxies lie in the most significant overdensities in the cosmic web. The stellar components of individual galaxies make up only a small fraction of a cluster's mass \citep[$f_* \sim 2-5\%$, e.g.,][]{2013ApJ...778...14G}, but they are relatively easily observed and trace the conditions of the cluster through cosmic history. In addition to being used to identify clusters \citep{2000AJ....120.2148G, 2007ApJ...660..239K, 2007ApJ...660..221K}, photometric observations of cluster galaxies have been used to constrain the processes that drive galaxy evolution \citep{1978ApJ...219...18B, 1984ApJ...285..426B, 1981ApJ...251L..75K, 1992ApJS...78....1D, 2007MNRAS.374..809D, 2008ApJ...673..742G, 2009MNRAS.399.1858L} and to characterize the properties of their entire host systems \citep{1974ApJ...194....1O, 2003ApJS..148..243B, 2003ApJ...585..215Y, 2007ApJ...669..905B, 2010MNRAS.404.1922A}.\\

\begin{figure*}
\centering
	\includegraphics[width=\textwidth]{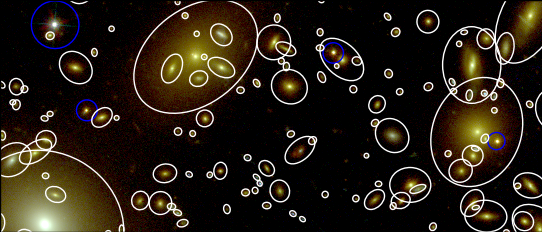}
    \caption{A portion of the MACS J0717 field, seen in F125W (red), F814W (green), and F555W (blue). Objects detected in this work are outlined in white ellipses corresponding to the region used for photometry. Also shown are objects classified as stars, which are marked by blue circles. This image is approximately $70''$ wide.}
\label{fig:sources_detected_m0717}
\end{figure*}

Since its launch in 1990, the {\it Hubble Space Telescope} ({\it HST}) has enabled high-quality photometric observations of clusters. Early works focused on single pointings, usually with one filter \citep[e.g.,][]{1994ApJ...430..121C, 1994ApJ...430..107D, 1997ApJ...474..561O}. Mosaicked exposures allowed more complete observations of individual clusters, such as CL 1358+62 \citep{1998ApJ...500..714V, 2000ApJ...531..137K} and MS 1054$-$03 \citep{1999ApJ...520L..95V, 2000ApJ...541...95V}. With the installation of the Advanced Camera for Surveys \citep[ACS,][]{1998SPIE.3356..234F}, more detailed surveys of clusters were begun, including the ACS Intermediate Redshift Survey \citep{2003ApJ...596L.143B, 2006ApJ...644...30B, 2004ASSL..319..459F, 2005ApJ...623..721P}, as well as deep surveys of the closest clusters, including the Virgo \citep{2004ApJS..153..223C}, Coma \citep{2010ApJS..191..143H}, and Fornax \citep{2007ApJS..169..213J} clusters. However, until recently, there has been no survey of a large number of clusters with comprehensive photometric coverage.\\

To study the evolution of faint cluster galaxies, we desire a survey that has deep imaging of a sample of clusters covering a range of redshifts. In addition, this ideal cluster survey would have comprehensive, multi-wavelength photometry, allowing for quality photometric redshift estimates and better analysis of stellar population parameters.  Increased metallicity, dust content, age, and redshift can all redden a galaxy, but these properties affect its spectrum in different ways. Multi-wavelength photometry is therefore needed to break degeneracies in stellar populations, which cannot be done with few-band colors \citep{1994ApJS...95..107W, 1999ASPC..192..283W}. \\

The Cluster Lensing and Supernova Survey with Hubble \citep[CLASH,][]{2012ApJS..199...25P} is a deep cluster survey with broad wavelength coverage; it has 16-band (UV to IR) observations of 25 massive galaxy clusters at redshifts $0.2 \lesssim z \lesssim 0.9$ obtained with the {\it HST}. We show in Figure \ref{fig:sources_detected_m0717} a sample region of one CLASH cluster. Because of the crowding and spatially variable backgrounds, detailed studies of cluster galaxy properties with CLASH, aside from the central brightest cluster galaxies \citep[BCGs,][]{2012ApJ...756..159P, 2015ApJ...805..177D, 2017ApJ...835..216D, 2015ApJ...813..117F}, have been limited due to the difficulty in obtaining reliable photometry \citep{2014A&A...562A..86J}. At this observational depth, the light from cluster galaxies is observed co-spatially with the intracluster light (ICL) and the extended wings of massive galaxies such as the BCGs; this effect is apparent in Figure \ref{fig:sources_detected_m0717}. Disentangling the flux contributions from individual galaxies and the scale- and wavelength-variant background light distribution is a significant challenge for utilizing not only the CLASH data set, but also any deep survey of clusters of galaxies.\\

Previous works have dealt with the complex light distribution of clusters by parametrically modeling large galaxies and the ICL \citep[e.g.,][]{2010ApJS..191..143H, 2012ApJ...756..159P, 2016A&A...590A..30M}. Here, we propose an alternative method: using the mode of nearby pixels on a range of physical scales to determine the distribution of background light for each pixel. The mode is a difficult quantity to determine for even moderately well sampled data, and so several approximations exist, such as those of \citet{Pearson343} and \citet{1996A&AS..117..393B}. These approximations are not robust, however, and are not suitable for precision photometry. We therefore propose a new technique to not only measure the modal value of a background distribution, but also to measure the overall shape of the distribution itself.\\

Through the use of our new modal estimation techniques, we model each galaxy's individual photometric background pixel-by-pixel, recovering the local structure in the backgrounds on the broad range of physical scales that can contaminate photometry, and thereby accurately measure the flux of galaxies in the CLASH clusters. The photometric catalog produced in this way traces the cluster population down to ${\sim} M^*+4.5$ for all clusters and down to ${\sim} M^* + 7$ for the closest ($z \sim 0.2$) clusters, as measured by the magnitude at which 90\% of the galaxies expected by our luminosity function are present. We compute photometric redshifts for the galaxies in these cluster fields; in comparison to previous results for the CLASH clusters, we reduce the median absolute deviation in offset between photometric and spectroscopic redshifts by ${\sim}30$\% and find significant improvement in the purity of our sample for cluster science. Finally, we fit spectral energy distributions (SEDs) to our photometry to provide a catalog of stellar properties for the galaxies in the CLASH fields. Throughout this work, we assume a $\Lambda$CDM cosmology, with $\Omega_{\rm M} = 0.3$, $\Omega_\Lambda = 0.7$, and ${\rm H}_0 = 70\ {\rm km}\ {\rm s}^{-1}\ {\rm Mpc}^{-1}$.\\

\section{Data Set}

\begin{figure*}
\centering
		\includegraphics[width=\textwidth]{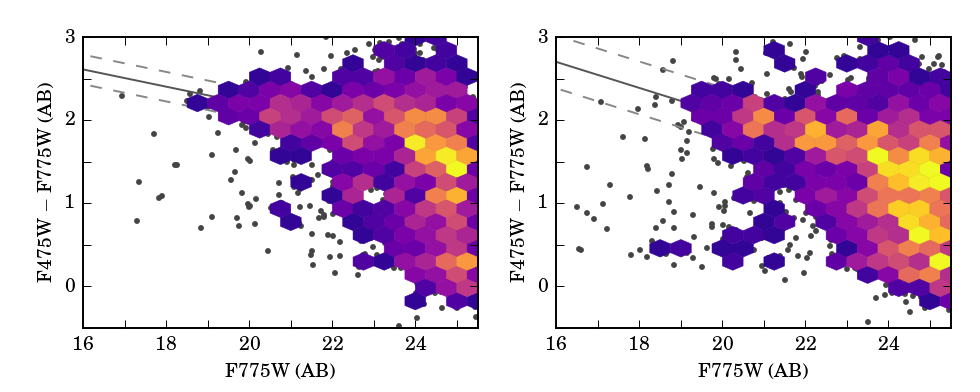}
    \caption{Observed F475W$-$F775W color-magnitude diagram for all galaxies detected in the fields of four CLASH clusters at redshift z$\approx 0.35$ from this work (left) and \citet[right]{2012ApJS..199...25P}. Also shown are linear fits to the red sequence (solid) with the intrinsic scatter (dashed); the intrinsic scatter with our photometry is ${\sim} 2\times$ smaller than with the original data. The improvement in colors is a result of the mode-measuring techniques described in this work. }
\label{fig:cmd_demonstration}
\end{figure*}

Imaging data in this work come from CLASH. This {\it HST} Multi-cycle Treasury Program imaged 25 galaxy cluster in 16 filters, covering the UV to the IR. CLASH clusters were chosen from the Massive Cluster Survey \citep[MACS,][]{2001ApJ...553..668E, 2007ApJ...661L..33E,2010MNRAS.407...83E} and the Abell catalog \citep{1958ApJS....3..211A,1989ApJS...70....1A}. Of the 25 clusters, 20 \citep[16 from][]{2008MNRAS.383..879A} were chosen due to their dynamically relaxed X-ray morphology and five chosen due to their strength as strong lenses. With an average of 20 orbits per cluster, the {\it HST} component of CLASH provides us with an unprecedented look into the environments of galaxy clusters. \\

The CLASH dataset combines observations from multiple instruments, which entails differing coverage for each filter. The ACS field-of-view is approximately $202'' \times 202''$ while the WFC-IR field covers roughly 40\% of that area, so many objects away from the cluster center lack IR coverage. {\it HST} filters used in this work have the naming convention that a filter of central wavelength NNN is labeled as either ``FNNNW'' or ``FNNNLP,'' depending on whether it is a wide or long-pass filter, respectively (see Figure 10 in \citealt{2012ApJS..199...25P} for the spectral responses of filters used in this work). For central wavelengths longer than 1 $\mu {\rm m}$, NNN is listed in hundredths of $\mu {\rm m}$; below 1 $\mu {\rm m}$, NNN is in nm (for these filters, NNN is always greater than 200). Filter properties are listed in Table \ref{tab:filter_properties}. For some clusters we also used archival F555W images. Throughout this paper we use the public full-depth CLASH {\it HST} mosaics\footnote{https://archive.stsci.edu/prepds/clash/}, which were produced using the approaches described in \citet{2011ApJS..197...36K} and references therein. We only use CLASH {\it HST} mosaics binned to $0\farcs065$ per pixel.\\

\begin{deluxetable}{lrrrr}
\tabletypesize{\scriptsize}
\tablecaption{Filter Parameters}
\tablewidth{0pt}
\tablehead{
\colhead{Filter Name} & \colhead{Instrument} & \colhead{${\rm zp}_{\rm AB}$} &  \colhead{$\lambda_{\rm pivot}$} & \colhead{$\Delta \lambda$}\\
\colhead{} & \colhead{} & \colhead{(mag)} & \colhead{(\AA )} & \colhead{(\AA )}}
\startdata
F225W & WFC3-UVIS & 24.0966 & 2359\tablenotemark{a} & 467\tablenotemark{a} \\
F275W & WFC3-UVIS & 24.1742 & 2704\tablenotemark{a} & 398\tablenotemark{a} \\
F336W & WFC3-UVIS & 24.6453 & 3355\tablenotemark{a} & 511\tablenotemark{a} \\
F390W & WFC3-UVIS & 25.3714 & 3921\tablenotemark{a} & 896\tablenotemark{a} \\
F435W & ACS-WFC & 25.6578 & 4328 & 1018\tablenotemark{b}\\
F475W & ACS-WFC & 26.0593 & 4747 & 1422\tablenotemark{b}\\
F555W & ACS-WFC & 25.7347 & 5361 & 1263\tablenotemark{b}\\
F606W & ACS-WFC & 26.4912 & 5921 & 2236\tablenotemark{b}\\
F625W & ACS-WFC & 25.9067 & 6311 & 1396\tablenotemark{b}\\
F775W & ACS-WFC & 25.6651 & 7692 & 1493\tablenotemark{b}\\
F814W & ACS-WFC & 25.9593 & 8057 & 2338\tablenotemark{b}\\
F850LP & ACS-WFC & 24.8425 & 9033 & 2063\tablenotemark{b}\\
F105W & WFC3-IR & 26.2707 & 10552\tablenotemark{a} & 2650\tablenotemark{a} \\
F110W & WFC3-IR & 26.8251 & 11534\tablenotemark{a} & 4430\tablenotemark{a}\\
F120W & WFC3-IR & 26.2474 & 12486\tablenotemark{a} & 2845\tablenotemark{a}\\
F140W & WFC3-IR & 26.4645 & 13923\tablenotemark{a} & 3840\tablenotemark{a}\\
F160W & WFC3-IR & 25.9559 & 15369\tablenotemark{a} & 2683\tablenotemark{a}
\enddata
\tablenotetext{a}{Taken from the WFC3 Instrument Handbook \citep{2016wfci.book.....D}.}
\tablenotetext{b}{95\% cumulative throughput width}
\label{tab:filter_properties}
\end{deluxetable}

As part of the public release of CLASH data, \citet{2012ApJS..199...25P} released a catalog of photometry for objects in all 25 CLASH filters, not optimized for cluster members. In Figure \ref{fig:cmd_demonstration} we show a color-magnitude diagram for all objects detected in the fields of four CLASH clusters at redshifts $\langle z \rangle = 0.350 \pm 0.005$: MACS 1115, MACS 1931, RXJ 1532, and RXJ 2248. Data from \citet{2012ApJS..199...25P} are shown on the right; data we will present in this work are shown on the left. Our new work not only has greater purity for cluster studies, but we find a red sequence\footnote{Red sequence fitting was performed using the \texttt{LTS\_LINEFIT} program described in \citet{2013MNRAS.432.1709C}, based on the methods of \citet{2006DMKD...12....1R}.} for these clusters with decreased scatter ($\sigma_{int} = 0.18$ mag with our data, compared to $\sigma_{int} = 0.32$ mag with the previous release). Our measured red sequence slope and scatter are more consistent with previous studies (slope: e.g., \citealt{1998ApJ...501..571G, 2007MNRAS.374..809D, 2009MNRAS.394.2098S, 2014MNRAS.441..776T}; scatter: e.g., \citealt{1992MNRAS.254..601B, 1998ApJ...492..461S, 2001MNRAS.326.1547T, 2005ApJ...619..193M, 2009ApJ...695.1058R}). As derived properties from the CLASH dataset -- photometric redshifts and SED fits -- are based on the set of colors in all 16 CLASH filters, our new work will provide us with the ability to reliably infer cluster galaxy properties in a way the original work did not.

\subsection{Statistical Background Light Estimators}

One of the fundamental assumptions of this work is that, for a pixel containing the light from a galaxy, the observed flux in that pixel is the sum of light from the galaxy and from the background light drawn from some unknown distribution. This background distribution may include contributions from physical structure (such as the ICL and other galaxies), contaminating light (such as star spikes and scattered light), and the sky (including instrumental effects). Lacking a complete understanding of the background light due to the limitations of our telescope's optics and the finite observation time, our best solution is to model the background light distribution from nearby pixels. Accurately describing the background brings three specific challenges: determining a nominal measure of the expected value of the background light, determining the range of that distribution, and performing this characterization with a limited sample of pixels, some of which may be outliers from a separate distribution (such as from the wings of a nearby galaxy)\\

Galaxy clusters are particularly challenging environments for background estimation. While a number of works have considered the problem of crowded-field photometry for stellar clusters, galaxies have significant angular extents, ensuring overlapping profiles and extreme difficulty in estimating local photometric background levels. And unlike stars, which are point sources described by their point spread function (PSF), galaxies are extended surface brightness distributions convolved with the PSF. Furthermore, clusters are filled with ICL, stellar emission associated with the cluster but not with any individual galaxy \citep{1999ASPC..170..349V,2005ApJ...631L..41M,2015MNRAS.448.1162D}. Accurately counting the flux of the ICL in an aperture can impact the observed colors of galaxies \citep{2005MNRAS.364.1069D, 2005MNRAS.358..949Z, 2007ApJ...656..756W, 2010ApJ...720..569R}, but improperly accounting for a galaxy's full extent due to the ICL can impact the total measured magnitude. Any cluster photometry routine must be able to deal with a spatially and chromatically varying background.\\

A number of routines have been introduced to deal with the complexities of cluster photometry, many of which utilize parametric modeling of cluster galaxies through tools such as {\tt GALFIT} \citep{2002AJ....124..266P, 2010AJ....139.2097P} and techniques including shapelets \citep{2005MNRAS.363..197M} and Chebyshev Rational Functions \citep[CHEFs,][]{2012ApJ...745..150J}.  \citet{2014ApJ...781...24G} simultaneously fit {\tt GALFIT} galaxy light profiles and a diffuse background model for CL0024+17. \citet{2016A&A...590A..30M} modeled a limited set of galaxies and the ICL independently for two Frontier Fields clusters, and then ran detection and photometry routines using the model-subtracted images. \citet{2017MNRAS.470...95M} modeled all of the galaxies in the CLASH fields with CHEFs, and subtracted off the residual light to compute background-free photometry. Maps of the Frontier Fields ICL were produced by \citet{2017ApJ...846..139M} by combining the measured background from postage stamp {\tt GALFIT} models pixel-by-pixel. \citet{2017ApJ...835..113L} mapped and subtracted ICL in the Frontier Fields through the use of wavelet decomposition. The common theme of these works is that background light is computed as the residual light after modeling galaxies. We have implemented a method that robustly and locally computes the mode over scales that capture the spatially variable background, which we use to calculate a pixel-by-pixel background on a galaxy-by-galaxy basis, without needing to model individual galaxies.\\

Estimating a common background level in a set of pixels has a storied history \citep{1980A&A....84...81B, 1987PASP...99..191S, 1989PASP..101..616H, 1992ASPC...23...90D, 1996A&AS..117..393B, 1998MNRAS.296..339N, 1999AJ....117.2757Z, 2011AJ....142...31B}. The simplest approach is to estimate the first moment, or mean, of the distribution, but this estimator is notoriously biased \citep[see the discussions in, e.g.,][]{rey1983introduction, 1990AJ....100...32B}. Beyond that, for regions containing multiple flux distributions (e.g., sky and a diffraction spike from a bright star), simple characterizations of the background will be inherently inaccurate. In this work, we treat the set of pixels as a distribution of distinct observations of the background, where some fraction of pixels may be contaminated by unknown interloping sources of flux. As such, the most frequent occurrence in the set, the mode, ought to represent an unbiased estimate of the photometric background. Each pixel also has an associated random measurement error, which we treat as symmetric about zero. We discuss below two techniques for estimating the mode: a computationally fast but less robust technique for detecting galaxies, and a more robust yet computationally intensive method to measure the peak and distribution of background pixels for photometry.\\
\subsection{Detection Images}

\begin{figure*}
\centering
	\begin{minipage}{0.3\textwidth}
		\centering
		\includegraphics[width=\textwidth]{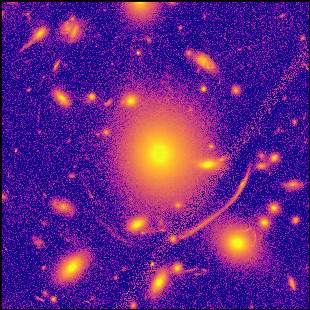}
		\vspace{0.0ex}
	\end{minipage}
	\begin{minipage}{0.3\textwidth}
		\centering
		\includegraphics[width=\textwidth]{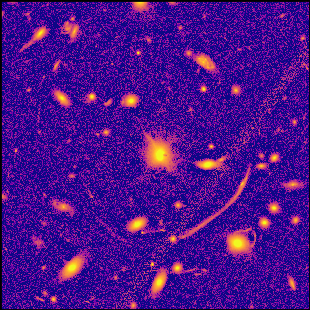}
		\vspace{0.0ex}
	\end{minipage}
	\begin{minipage}{0.3\textwidth}
		\centering
		\includegraphics[width=\textwidth]{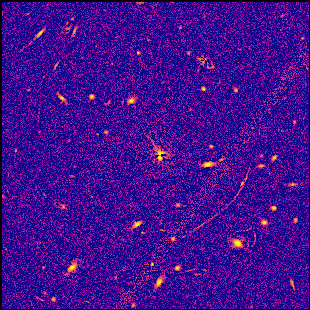}
		\vspace{0.0ex}
	\end{minipage}
    
    	\begin{minipage}{0.3\textwidth}
		\includegraphics[width=\textwidth]{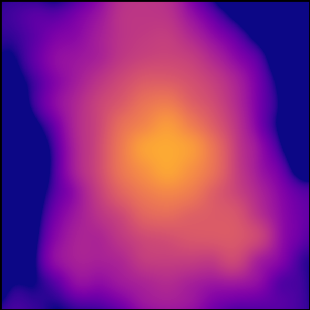}
    \end{minipage}
	\begin{minipage}{0.3\textwidth}
		\includegraphics[width=\textwidth]{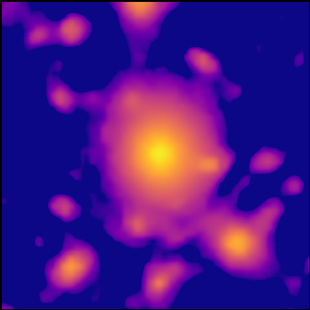}
    \end{minipage}
	\begin{minipage}{0.3\textwidth}
		\includegraphics[width=\textwidth]{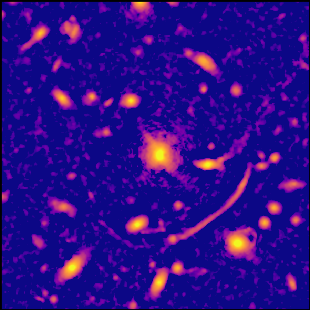}
    \end{minipage}

    \caption{{\bf Top:} residual structure in the central region of Abell 383 after subtracting backgrounds computed with the method of steepest ascent on scales of 128, 16, and 4 pixels ($8\farcs32$, $1\farcs04$, and $0\farcs26$), respectively.
{\bf Bottom}: large, medium, and small-scale structure of Abell 383, computed by taking the difference of background-subtracted images with backgrounds computed at scales of 256 and 128, 128 and 16, and 16 and 4 pixels, respectively. The suite of images is used to detect small galaxies using the images on the  right, and the full extent of large galaxies using the images on the left. }
\label{fig:mode_detection_images}
\end{figure*}

Detecting objects with angular size scales ranging over two orders of magnitude requires a multiresolution analysis of the images. To identify small galaxies, we need an image where small-scale structure is visible, while to identify large galaxies we require a map of the large-scale structure. We create these images by using the method of steepest ascent (described below) to measure the mode in a sliding box across each image, where we use different box sizes to create unbiased maps of the spatially varying background over the broad range of scales needed to detect both small and large objects. In this work, we use boxes with sizes in seven scales separated logarithmically between 4 and 256 pixels ($0\farcs26$ and $16\farcs64$), inclusive. Three of the background-subtracted images for Abell 383 are shown in the top panel of Figure \ref{fig:mode_detection_images}.\\

\begin{figure*}
\centering
	\includegraphics[width=\textwidth]{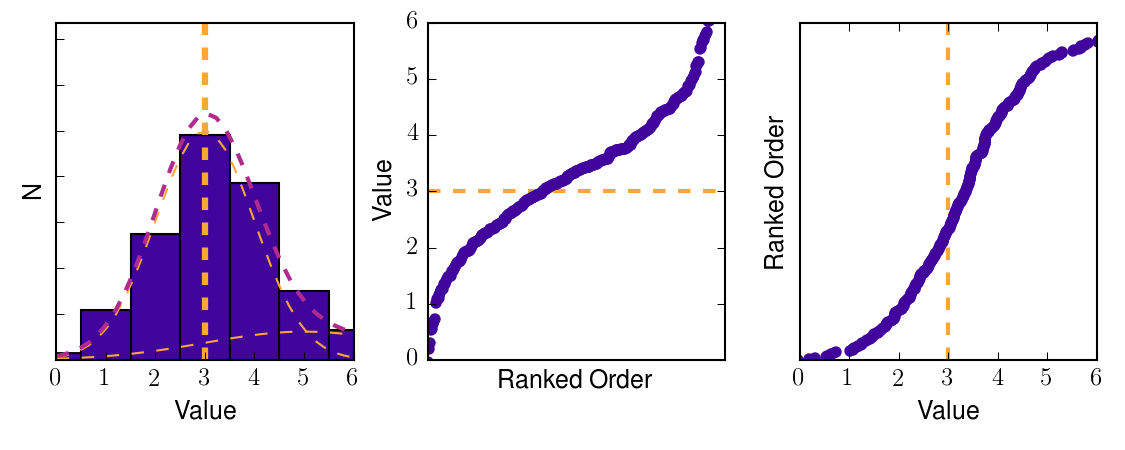}
    \caption{Schematic representation of the method of steepest ascent for finding the mode. On the left, we present a sample of points drawn from two gaussian distributions (orange); the peak of this distribution is the mode, marked by a vertical orange line. By plotting the measured value as a function of ranked order (middle), we see that the inflection point corresponds to the mode (orange). The transpose of this (right), which shows the cumulative count of each value, is modeled by a cubic spline.}
\label{fig:steepest_ascent_explanation}
\end{figure*}

To create the background-subtracted maps, we determine a local background for each pixel by using the method of steepest ascent to calculate the mode of the set of background pixel values (this technique has also been adapted by \citealt{2017ApJ...839..127P}). The mode corresponds to the value at which most sample points are expected to lie; in other words, it is the flux with the highest density of points in the sample. To determine this value, we first sort each pixel flux in a sample of $N$ pixels from smallest to largest, ${\rm F}_1\leq {\rm F}_2 \leq ... \leq {\rm F}_{{\rm N}-1} \leq {\rm F}_{\rm N}$. For a well-sampled distribution, this ordered set will have a range of values such that $F_{i + j} - F_{i - j} \approx 0$, where $i$ is the index of the mode and $j$ is a sampling range. The transpose of this distribution is the number of points with flux less than $f$, ${\rm N}_{f \leq 0} \leq {\rm N}_{f \leq i} \leq ... \leq {\rm N}_{f \leq F-i} \leq {\rm N}_{f \leq F}$. In this distribution, the mode is therefore the region of steepest ascent, where ${\rm N}_{f \leq i + j} - {\rm N}_{f \leq i - j}$ is maximized. To calculate the point of steepest ascent, we fit the transpose function with a cubic spline, so that the steepest ascent is calculated numerically. This technique is shown schematically in Figure \ref{fig:steepest_ascent_explanation}. While this technique allows for rapid computation of the mode, it is limited in its accuracy in the presence of complex backgrounds (which may distort a polynomial fit) and for small samples. Due to these concerns, we use this method for computational expediency in estimating the background for purposes of identification and detection of sources, but not for flux estimation. Since our science does not depend on teasing out detections of the faintest, smallest-background sources, small errors in the mode for individual pixels do not affect detection and source definition for our purposes here.\\

One issue that hampers detection of objects in cluster fields is source confusion, whereby two nearby galaxies are detected as one, creating a pseudo-object that includes both of them but is concentric with neither. To mitigate this issue, we produce a separate suite of detection images (shown in the bottom panel of Figure \ref{fig:mode_detection_images}) from the previously generated background-subtracted images (shown in the top panel of Figure \ref{fig:mode_detection_images}); here, the new detection image is the difference of the background-subtracted images at neighboring scales (e.g., the image with an 8 pixel background scale was subtracted from that with a 16 pixel background scale). In essence, we are performing an {\it \'{a} trous} wavelet transform \citep[such as used by][]{2017ApJ...835..113L} on the original images, but with the critical difference that the dual is not uniform across the field due to the nonlinear nature in which the mode is calculated. This strategy removes smaller sources from the detection areas of larger galaxies.\\

Our method of determining the mode does not perform well at small scales when no clear background can be determined, such as at the center of a massive galaxy.  In this regime, the background selection region does not include any background pixels, so the background subtraction routine subtracts real signal. The effect of performing background subtraction without having any background can be seen in the upper right panel of Figure \ref{fig:mode_detection_images}. To properly photometer these galaxies, we need to be able to subtract a background in their central regions consistent with the background at the outer regions. Because the generated background-subtracted images lack that flexibility, we only use these images for detection and not for photometry.\\

\subsection{Source Detection}
\label{sect:Detection}

\begin{deluxetable*}{lccccc}
\tabletypesize{\scriptsize}
\tablecaption{Source Extractor Detection Parameters}
\tablewidth{0pt}
\tablehead{
\colhead{Detection\tablenotemark{a}} & \colhead{\texttt{CLEAN\_PARAM}\tablenotemark{b}} & \colhead{\texttt{DEBLEND\_MINCONT}\tablenotemark{b}} & \colhead{\texttt{DEBLEND\_NTHRESH}\tablenotemark{b}} & \colhead{\texttt{DETECT\_MINAREA}\tablenotemark{b}} & \colhead{Max Offset} \\  
\colhead{ } & \colhead{ } & \colhead{ } & \colhead{ } & \colhead{(pix)} & \colhead{(pix)} }
\startdata
004              & 0.2 & 0.40 & 50 & 15 & \nodata \\
008$-$004 & 0.2 & 0.20 & 60 & 16 & 6\\
016$-$008 & 0.3 & 0.10 & 60 & 18 & 6\\
032$-$016 & 0.4 & 0.10 & 60 & 20 & 8\\
064$-$032 & 0.5 & 0.10 & 60 & 20 & 8\\
128$-$064 & 0.6 & 0.10 & 60 & 20 & 12\\
256$-$128 & 0.2 & 0.10 & 60 & 20 & 12\\
Stars             & 0.4 & 0.10 & 60 & 20 & \nodata 
\enddata 
\tablenotetext{a}{Numerical detection images are either the background size of mode filtering (004) or the backgound sizes of the subtracted images (e.g., 008$-$004)}
\tablenotetext{b}{\texttt{CLEAN\_PARAM} is the efficiency of cleaning artifacts of bright sources from the detection list; a lower value of \texttt{CLEAN\_PARAM} fits more extended structure to bright sources, resulting in a more aggressive cleaning. \texttt{DEBLEND\_MINCONT} and \texttt{DEBLEND\_NTHRESH} determine whether Source Extractor separates detections into multiple objects. For \texttt{DEBLEND\_NTHRESH} logarithmically spaced flux bins, objects are deblended into two objects if they both have \texttt{DEBLEND\_MINCONT} of the flux of their combined flux measure. \texttt{DETECT\_MINAREA} is the minimum required number of pixels above the detection threshold for a galaxy to be detected by Source Extractor.}
\label{tab:source_extractor_parameters}
\end{deluxetable*}

We have nonparametric models of backgrounds mapped on scales from 8 to 256 pixels, as well as images with subtracted backgrounds of 4 pixels, for the non-UVIS filters for each cluster. We run Source Extractor \citep{1996A&AS..117..393B} on each, with parameters listed in Table \ref{tab:source_extractor_parameters}. We use an RMS map based on the weight map produced by MosaicDrizzle for each filter and cluster combination, which is itself an inverse variance image based on the input exposures that contributed to each pixel. For each detection, Source Extractor outputs the geometric properties of the detection ellipse, specifically WCS and pixel coordinates, semi-major and -minor axes, position angle, and the \texttt{KRON\_RADIUS}\footnote{The Kron radius \citep{1980ApJS...43..305K} is a radius selected to capture more than 90\% of a galaxy's flux. See \cite{2005PASA...22..118G} for a discussion of its usage in Source Extractor.}.\\

We assemble a detection catalog for each individual filter by working from the detection images with the largest background regions to the smallest (from left to right in Figure \ref{fig:mode_detection_images}). For every object detected in one image, we check to see if there was a match in the next-smallest detection image within a small offset, as specified in Table \ref{tab:source_extractor_parameters}. Those that had matches, as well as unmatched objects from the smaller catalog, are passed on to the next scale. For all but the smallest scale, the actual detection is performed on a subtraction image (the bottom row of Figure \ref{fig:mode_detection_images}); therefore, our technique has the result of detecting the full extent of galaxies and propagating those sizes down to the small-background images.\\

After collecting all the detections from each filter into one catalog, we then create a master source list for each cluster field based on the multi-wavelength detection suite. We merge the detection catalogs filter-by-filter, combining objects with peak flux values located near each other. As the geometric properties differed between detections in different filters, we reduce each object to the properties from one filter. These properties are from the detection with the fourth-largest size -- or, for objects detected in fewer than eight filters, the median size -- where the size was the sum of semi-major and minor axes. The choice to not use the largest detection of an object was motivated by source confusion lumping multiple objects into excessively large apertures; the choice of fourth-largest detection was motivated by experimentation. When objects were detected in multiple filters, the aperture sizes would quickly converge to the same size, but several detections could be excessively large if multiple nearby structures blended together at one background scale. After trying multiple methods of selecting an appropriately sized aperture from the detections in every filter, we found that skipping the three largest detections would remove the unphysically large apertures while not causing measured galaxy sizes to otherwise shrink (as aperture sizes converge toward the same value). We trim this catalog to only those objects detected in at least three filters. To avoid detecting diffraction spikes from stars, we also only include those objects with semimajor axis no more than eight times the length of the semi-minor axis or with a semi-minor axis of at least 5 pixels, which excludes long, thin detections without removing larger elliptical objects, such as edge-on spirals.\\

To remove stars from our catalog, we run Source Extractor on the original images for each filter, using the parameters given in Table \ref{tab:source_extractor_parameters}. Detections in each filter are combined in the same manner as before; to only include stars, we exclude any object with a \texttt{CLASS\_STAR} value below 0.9. We match this star catalog with our previous detection catalog, and those objects included in both are marked as stars. We generate a circular mask aperture with a radius equal to the semi-minor axis from our original detection image for each star.\\

\begin{deluxetable}{llll}
\tabletypesize{\scriptsize}
\tablecaption{Star Masks, $r \geq 1\farcs0$}
\tablewidth{0pt}
\tablehead{
\colhead{$\alpha_{2000}$} & \colhead{$\delta_{2000}$} &  \colhead{$r$} & \colhead{Cluster}\\
\colhead{} & \colhead{} & \colhead{($''$)} & \colhead{}}
\startdata
01:31:57.74 & $-$13:34:43.5 & 1.26 & Abell 209 \\
01:31:53.94 & $-$13:35:58.1 & 3.42 & Abell 209 \\
01:31:50.73 & $-$13:37:23.2 & 1.39 & Abell 209 \\
01:31:47.74 & $-$13:37:24.0 & 1.47 & Abell 209 \\
02:47:55.00 & $-$03:30:41.4 & 2.46 & Abell 383 \\
02:48:05.48 & $-$03:30:59.0 & 2.21 & Abell 383 \\
02:48:05.69 & $-$03:31:16.9 & 1.50 & Abell 383 \\
02:48:09.28 & $-$03:31:32.9 & 2.94 & Abell 383 \\
02:48:02.83 & $-$03:31:32.8 & 1.08 & Abell 383 \\
02:48:00.14 & $-$03:31:34.7 & 2.03 & Abell 383 \\
\enddata
\tablecomments{This table is available in its entirety in machine-readable form.}
\label{tab:clash_star_coords}
\end{deluxetable}
After creating the master detection images for each cluster, we inspect them by eye. We verify the accuracy of our star masks, reclassifying easily identifiable objects that our pipeline had misidentified as stars (an average of four objects per field were reclassified). To ensure repeatability, we provide the coordinates and radii of the star masks used in this work with mask radii of $r > 2''$ in Table \ref{tab:clash_star_coords}. Second, we clean the detection catalog of over-detections (such as in a spiral galaxy being detected at both the galaxy level and as individual knots) and adjust a small number of mis-proportioned ellipses to circles with size given by either their semimajor or semiminor axes (usually caused by extended structure from other, nearby galaxies). We also had to create a new aperture for the BCG of Abell 2261, which has an unusually flat central surface brightness profile \citep{2012ApJ...756..159P}, making it difficult for our technique to link large-scale detections to a central point at small background scales. As an example of our detection efficiency, the detection regions for a section of MACS J0717 are shown in Figure \ref{fig:sources_detected_m0717}.\\

As with the creation of any photometric catalog, we tailored our detection strategy toward a specific science goal; in this case, the detection and photometry of cluster galaxies. Because of this decision, we did not prioritize the detection of strongly lensed arcs \citep[previously studied in the CLASH clusters by][]{2016ApJ...817...85X} or lensed galaxies \citep[e.g.,][]{2012ApJ...747L...9Z, 2013ApJ...762...32C}. By requiring detections in at least three filters, we minimized false detections, but also excluded drop-out galaxies at too large of a redshift to be seen in many filters. We also linked emission at multiple angular scales through their shared centers, making our technique unsuited for the complex morphologies of strongly lensed arcs.\\

\section{Photometry}

The primary challenge of photometring crowded fields is assigning the entirety of the observed flux across individual sources and the background. Here, we describe a new technique for accurately accounting for the flux of objects in a cluster field. Starting with the smallest galaxies, we measure a background at the outer edges of each object and work our way inward, replacing pixels with their measured background as we go. This replacement allows us to not only provide a background measurement at the inner parts of larger galaxies, but also to effectively peel off smaller galaxies from larger ones, letting us distribute the flux between overlapping objects accurately. As part of this measurement, we characterize the distribution of background light, which in turn gives us a per-pixel estimate of the expected variation in the background light; this preserves the noise properties of the background. We detail this process below.\\

We use the method of steepest ascent to determine the mode to estimate and subtract a background for source detection; this technique calculates a mode using individual flux measurements of background pixels. One additional input to consider when determining the background is the contribution of measurement uncertainty. Each pixel has both a flux and a flux uncertainty; this can be thought of as each pixel being a normalized probability distribution of fluxes. In this scheme, the mode will correspond to the flux value with the most total probability from the combined probability distributions of all the pixels in the background. Specifically, this means that the mode is the peak of a probability density function, which we estimate through a technique similar to a kernel density estimation, although where each pixel has a unique kernel. \\

Each background pixel is defined by two values: $f_i$ and $\sigma_i$, the flux and uncertainty, respectively. By convolving the flux in each pixel in the background region, $B$, with a Gaussian kernel of width $\sigma_i$, we create a probability distribution for the fluxes of all the pixels in the background region. The mode is then found as the peak of this distribution, which is given by
\begin{equation}\label{eqn:gaussian_sum}
P(x) = \sum_{i \in {\rm B}} \frac{1}{\sigma_i}\ e^{ - (x - f_i)^2 / (2 \sigma_i^2)}.
\end{equation}
To find the maximum of this function in a computationally expedient manner over a large number of points, we identify the zeros of its derivative, given by
\begin{equation}\label{eqn:gaussian_deriv_sum}
dP(x)/dx = \sum_{i \in {\rm B}} \frac{ (f_i - x)}{\sigma_i^3}\ e^{ - (x - f_i)^2 / (2 \sigma_i^2)}.
\end{equation}
Due to the limits of computational efficiency, we only employ the more accurate kernel density estimation using Equations \ref{eqn:gaussian_sum} and \ref{eqn:gaussian_deriv_sum} when measuring flux.\\

For a given pixel, the measured flux, $f_i$, is itself drawn from a probability distribution; the drizzling step combines multiple samples of that probability distribution to calculate a measurement and uncertainty for that pixel. Thus, $f_i$ has already been convolved with the noise before we convolve it with a gaussian kernel. We performed numerical simulations to analyze how well our technique could recover the original distribution from a sample of $N$ points averaged from $X$ draws (analogous to $N$ pixels made from $X$ exposures). For a normal distribution, finding the mode through our technique is an order of magnitude more precise at finding the central moment of the original distribution than through using the median or mean. For large $X$ ($X \gtrsim 25$) the drizzling step induces kurtosis, causing the estimate of $\sigma$ to be biased high, but our mode performs better at this test than the median or mean for more reasonable values of $X$. For skew normal probability distributions, the average of $X$ random draws (pixel values after drizzling) is offset from the peak of the skew normal distribution, so the recovered mode is systematically offset from the peak of original probability. The severity of this offset, which is toward the average of the probability distribution, increases with increasing $X$. Again, for the skew normal distribution, the mode is the most precise measurement. Our simulations show that, while drizzling leaves an imprint on the noise characteristics of a region, our mode technique can still precisely recover a distribution's central value, particularly for small numbers of drizzled exposures. \\ 

\begin{figure*}
\centering
	\begin{minipage}{0.3\textwidth}
		\includegraphics[width=\textwidth]{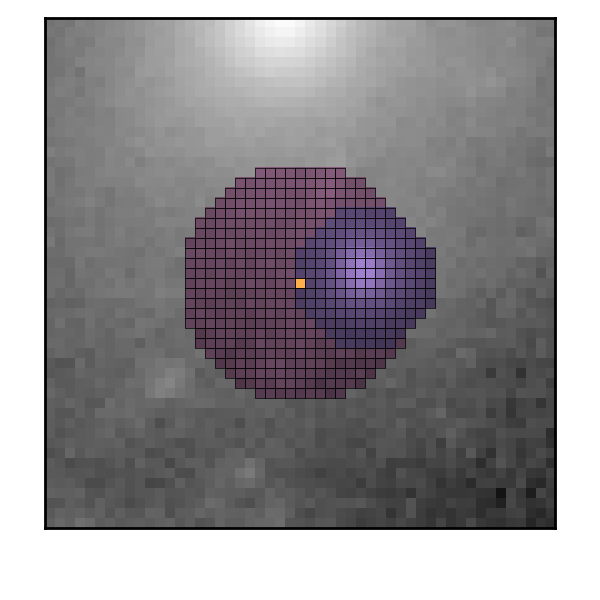}
    \end{minipage}
	\begin{minipage}{0.3\textwidth}
		\includegraphics[width=\textwidth]{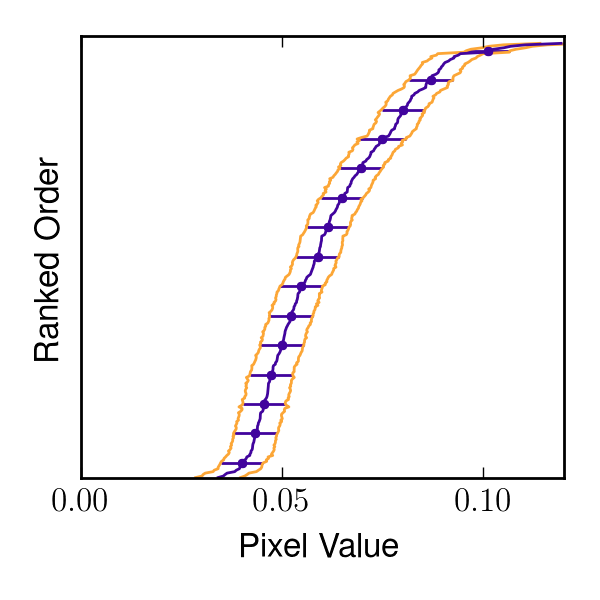}
    \end{minipage}
	\begin{minipage}{0.3\textwidth}
		\includegraphics[width=\textwidth]{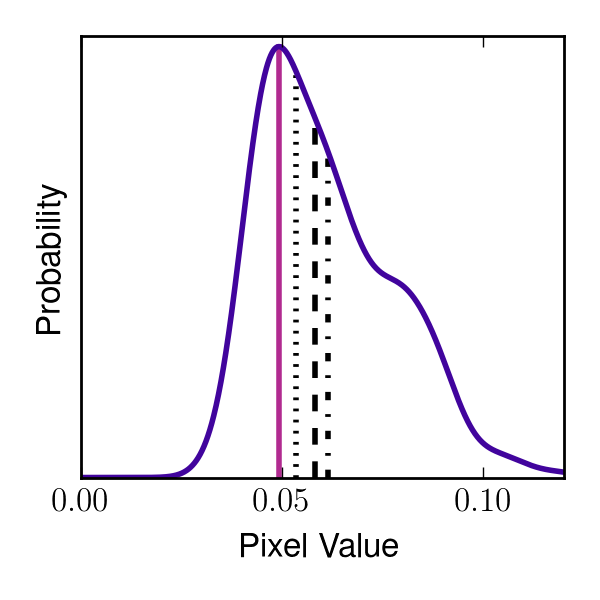}
    \end{minipage}
    \caption{Schematic representation of the photometry technique. On the left, we show an example of how the background for a pixel in a galaxy is selected. A background region (light red) is created around a pixel of interest (orange), avoiding any pixels in the aperture of the galaxy (purple). The middle panel shows the distribution of flux in those pixels; for the set of pixels ordered by flux, orange lines trace the uncertainties and the nominal values are shown in purple. For legibility, errorbars are only shown for every 20th pixel. Each of those flux measurements is convolved with a Gaussian kernel with $\sigma$ given by the pixel's variance from the weight map; the sum of those kernels is shown in purple in the right panel. The peak of this distribution (pink) is the background value for this pixel. In contrast, the mode found using the technique of \citet{1996A&AS..117..393B}, the median, and the mean are shown with dotted, dashed, and dotted-dashed lines, respectively. The uncertainty in this background determination is set by the flux values at which the probability has dropped to $e^{-1/2}$ times the maximum. The excess flux above this background for the pixel of interest is assigned to the galaxy, the flux of that pixel is set to the determined background value, and the uncertainty in the background is used to set a new value in the weight image for that pixel. To photometer the entire galaxy, this process is repeated for each pixel, working from the outer boundaries in. The background region in this example is slightly enlarged for demonstrative purposes.}
\label{fig:depeche_explained}
\end{figure*}

Photometry is performed on each galaxy, beginning with the smallest and working to the largest (as ranked by semi-major axis), on a pixel-by-pixel basis. Our technique involves eroding each galaxy from the outside in, replacing pixels with the value of the background around them, thereby propagating the conditions outside of the galaxy inward. For small galaxies overlapping with larger galaxies, this local background replacement preserves the structure of the larger galaxy. A schematic representation of how a background is measured is shown in Figure \ref{fig:depeche_explained}; the measured background in this example would then replace the value of the pixel highlighted in orange, and the photometry routine would advance to the next pixel. The width of the distribution of background pixels is propagated into the variance map, maintaining the statistical properties of the measured background. Our implementation of this technique for the CLASH observations is presented below.\\

When photometering each galaxy, we create an elliptical aperture based on the Source Extractor detection parameters. This region is blocky; pixels are either in the aperture or out of it, with no partial associations. We then assign an order of photometry by taking a one-pixel-wide annulus with outer radius equal to the galaxy's semi-major axis, and finding all of the galaxy pixels inside that ring. We shrink the annular radius pixel-by-pixel, noting the order of galaxy pixels to fall within it, until we have reached the centermost pixel. We then photometer the galaxy following that order.\\

To measure the flux in a pixel, we must first compute the background value. We consider a circular aperture around that pixel, with radius equal to $1.5 \times b$, where $b$ is the semi-minor axis of the galaxy. This value is constrained to lie within 3 and 12 pixels ($0\farcs195$ and $0\farcs78$, respectively), the former to ensure enough background pixels can be found and the latter to keep the background local to the galaxy. We exclude from this aperture any pixels contained within the galaxy itself that have not yet been photometered. For the remaining pixels, we pass their measured fluxes and uncertainties to our background measuring routine.\\

This routine, shown in Figure \ref{fig:depeche_explained}, convolves each flux measurement with a Gaussian kernel that has a standard deviation given by that flux measurement's uncertainty (drawn from the variance map of the image, which is itself updated for pixels that are replaced with background measurements). These distributions are then summed to create a single probability frequency curve for the background flux (as given in Equation \ref{eqn:gaussian_sum} and shown on the right of Figure \ref{fig:depeche_explained}). The nominal location of the background intensity is determined by finding the peak of this distribution; to do this, we employ a root finder on the derivative (given in Equation \ref{eqn:gaussian_deriv_sum}) to find all maxima. As these distributions can be multi-modal, it is important to find all maxima. We therefore step through the ordered range of flux measurements. If the sign of the derivative changes from positive to negative, we use these bounds to find a root. If the derivative is zero at any flux measurement, we add it to the list of roots. We ignore any changes from negative derivative to positive, as those mark minima. One implicit assumption is that maxima cannot occur bounded by two flux measurements at which the derivatives are the same sign -- these maxima would not be found by our root finder, which requires boundary values of opposite sign to function.\\

We use the interval bisection method of \citet{Brent:1973:AMD} as implemented in the \texttt{scipy} \texttt{brentq} algorithim to find the roots, given that the maxima are bounded. In the event of an error in this process, our code will find the mode using the Newton--Raphson method; as this method is unbounded, we seed it with an initial guess of the median of the background flux distribution. In both cases, the uncertainty on the background value is found by finding the flux value in both directions at which the probability distribution given by Equation \ref{eqn:gaussian_sum} is equal to $e^{-1/2} \times P( \mu )$, where $P( \mu )$ is the probability at the measured mode; this value is the relative height of a Gaussian distribution at $1 \sigma$. It is important to note that we are not looking for a quantity equivalent to the standard error in the mean, which would be inversely proportional to the square root of the number of pixels in the background; that quantity would be a measure of the accuracy of the mode. Instead, we characterize the intrinsic spread in the background, as each pixel has an additive flux component drawn from this distribution, and even with perfect knowledge of the background distribution, we cannot know the background flux for a certain pixel better than this level.\\

\begin{figure*}
\centering
	\begin{minipage}{0.3\textwidth}
		\includegraphics[width=\textwidth]{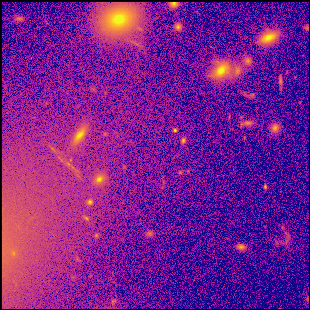}
    \end{minipage}
	\begin{minipage}{0.3\textwidth}
		\includegraphics[width=\textwidth]{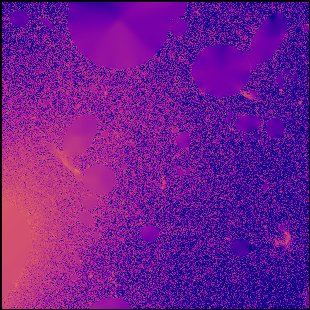}
    \end{minipage}
	\begin{minipage}{0.3\textwidth}
		\includegraphics[width=\textwidth]{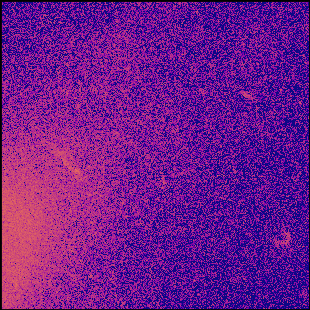}
    \end{minipage}
    \caption{Effect of our background modeling and subtraction technique for Abell 209. Images are from the F814W filter: the original (left), after every galaxy has been photometered (center), and after the subtracted image has been adjusted by a random draw from the final uncertainty model (right).}
\label{fig:depeche_results}
\end{figure*}

Having determined the background value for a pixel, the excess flux is assigned to the galaxy, while the pixel value is replaced with the background value and the width of the background probability distribution is incorporated into the variance map. We work from the outside of the galaxy in, so that the background measurement for the center of each galaxy is based on the backgrounds measured all around it. After the entire field has been photometered, we produce an image made by adjusting each pixel's flux by a random draw from the variance map; this step maintains the noise properties of the galaxy-subtracted image. We show the before, after, and randomly drawn images of a section of Abell 209 in Figure \ref{fig:depeche_results}.\\

In the galaxy-subtracted images of Figure \ref{fig:depeche_results}, the ICL is clearly visible. As \citet{2015MNRAS.448.1162D} and \citet{2017arXiv171011313D} have already performed a careful analysis of the ICL in CLASH clusters, we will not be discussing the residual ICL in this work, although the techniques presented here do offer a new route for studying ICL. Also visible in these images are some residual structures; as discussed in Section \ref{sect:Detection}, we did not prioritize detecting strongly lensed arcs. Although we do not obtain photometry for a complete sample of lensed objects, our photometric technique was designed such that undetected structures such as these arcs would not affect the background measurements.\\

\begin{deluxetable}{crrr}
\tabletypesize{\scriptsize}
\tablecaption{Photometric Properties}
\tablewidth{0pt}
\tablehead{ \colhead{Cluster} & \colhead{Filter} & \colhead{${\rm t}_{\rm exp}$} & \colhead{${\rm A}_{\lambda}$}\\
\colhead{} & \colhead{} & \colhead{(s)} & \colhead{(mag)} }
\startdata
Abell 209 & F225W & 7316.0 & 0.135 \\
Abell 209 & F275W & 7464.0 & 0.106 \\
Abell 209 & F336W & 4752.0 & 0.086 \\
Abell 209 & F390W & 4894.0 & 0.075 \\
Abell 209 & F435W & 4136.0 & 0.070 \\
Abell 209 & F475W & 4128.0 & 0.063 \\
Abell 209 & F606W & 4096.0 & 0.048 \\
Abell 209 & F625W & 4066.0 & 0.043 \\
Abell 209 & F775W & 4126.0 & 0.032 \\
Abell 209 & F814W & 8080.0 & 0.030 \\
\enddata
\tablecomments{This table is available in its entirety in machine-readable form.}

\label{tab:clash_photometric_properties}
\end{deluxetable}

We provide the photometric parameters used for each cluster and filter in Table \ref{tab:clash_photometric_properties}. As each processed image was reduced to a count rate, exposure times double as gain values. We assume Galactic extinction values taken from the NASA/IPAC Extragalactic Database (NED)\footnote{NED is operated by the Jet Propulsion Laboratory, California Institute of Technology, under contract with the National Aeronautics and Space Administration.}, which uses the \cite{2011ApJ...737..103S} recalibration of the \cite{1998ApJ...500..525S} infrared-based dust maps utilizing a \cite{1999PASP..111...63F} reddening law with $\textrm{R}_{\rm v} = 3.1$.\\

\subsection{Photometric Redshifts}
\label{sect:phot_redshifts_described}

We derive redshift probability distribution functions for each galaxy we observed using the Bayesian photometric redshift code presented by \citet{2000ApJ...536..571B} \citep[BPZ\footnote{http://www.stsci.edu/~dcoe/BPZ/}, ][]{2004ApJS..150....1B, 2006AJ....132..926C}. BPZ fits the broadband flux measurements with a set of empirical templates in a grid of redshifts and computes a probability distribtion function from those fits. This code was used to produce estimates of photometric redshifts for the default CLASH source catalog \citep{2012ApJS..199...25P}. \\

Our determination of $P(z)$ for each galaxy covers the range of redshifts from $z = 0.01$ to $z = 12.0$ in steps of $\Delta_z = 0.001$. We use 11 template galaxy spectra, including both early-type and late-type galaxy spectral templates, and we interpolate between these 11 to make an additional 90 templates. To account for zero-point uncertainties, we set a minimum photometric uncertainty of 0.02 mag \citep{2014PASP..126..711B, 2016AJ....152...60B}. \\

To characterize the accuracy of our redshifts, we compare a sample of galaxies with spectroscopically derived redshifts to our values. These values come from the CLASH-VLT collaboration \citep{2013A&A...558A...1B, 2016ApJS..224...33B, 2017MNRAS.466.4094M}, the Sloan Digital Sky Survey Data Release 13 \citep{2016arXiv160802013S}, as well as works by \cite{2002ApJ...573..524C}, \cite{2008MNRAS.387.1374M}, \cite{2009A&A...499..357G}, \cite{2009ApJ...693..617H}, \cite{2010MNRAS.404..325R}, \cite{2010ApJS..188..280S}, \cite{2012ApJ...757...22C}, \cite{2012AJ....144...79G}, \cite{2013ApJ...767...15R}, \cite{2014ApJS..211...21E}, and \cite{2014ApJ...783...52G}. Additionally, we use a sample of unpublished VLT-VIMOS redshifts for four cluster fields (P. Rosati \& M. Nonino 2017, private communication) and a sample of unpublished redshifts for nine clusters from the IMACS-GISMO instrument on Magellan (D. D. Kelson 2017, private communication).\\

\begin{figure*}
		\includegraphics[width=\textwidth]{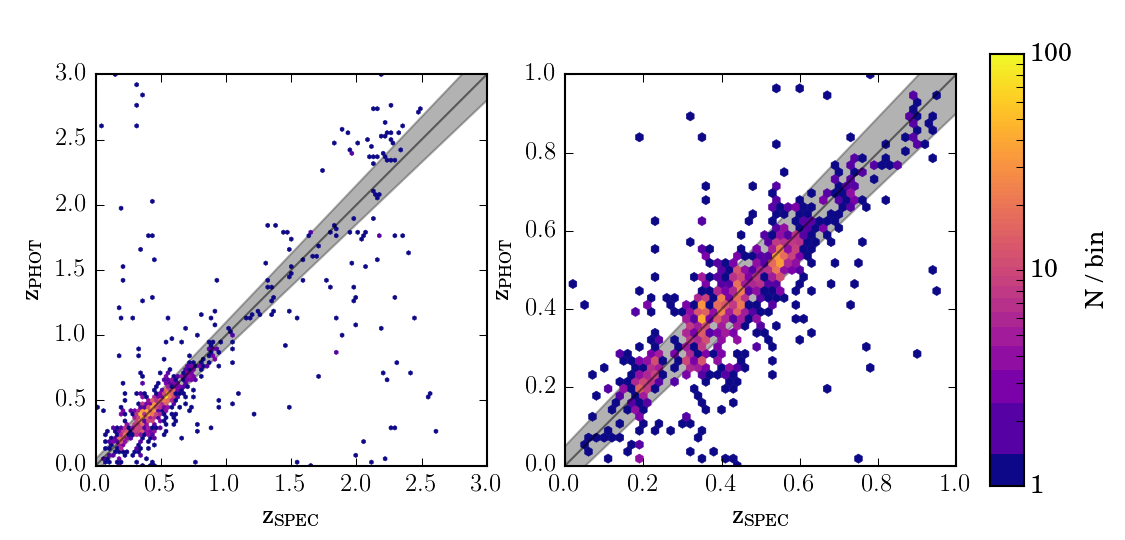}
    \caption{Comparison of photometric redshifts measured in this work to spectroscopic redshifts. The thin band traces the region where $| (z_{\rm p} - z_{\rm s}) / (1 + z)| < 0.05$. The full redshift coverage from $z=0$ to $z=3.0$ is shown on the left panel; a zoom-in to just $z=0$ to $z=1.0$ is shown on the right. Points are binned into hexagons, with the total density of points scaled logarithmically from 1 to 100 counts per hex, as indicated by the colorbar.}
\label{fig:spec_to_phot_comp}
\end{figure*}

\begin{deluxetable*}{lrr}
\tabletypesize{\scriptsize}
\tablecaption{Photometric Redshift Accuracy and Sample Purity}
\tablewidth{0pt}
\tablehead{ \colhead{ } & \colhead{This Work} & \colhead{\citet{2012ApJS..199...25P}\tablenotemark{a} }}
\startdata
Spectroscopic Matches & 1716 & 1942 \\
Well Matched\tablenotemark{b} & 1149 (66.96\%) & 1136 (58.50\%) \\
Minor Outliers\tablenotemark{c} & 325 (18.94\%) & 409 (21.06\%) \\
Substantial Outliers\tablenotemark{d} & 108 (6.29\%)  & 170 (8.75\%) \\
Catastrophic Outliers\tablenotemark{e} & 134 (7.80\%) & 227 (11.69\%) \\
Median Absolute Deviation\tablenotemark{f}, All Galaxies & 0.030 & 0.038\\
75th Percentile\tablenotemark{f}, All Galaxies & 0.053 & 0.073 \\
Median Absolute Deviation\tablenotemark{f}, $z_{\rm s} \leq 1.0$ & 0.027 & 0.035\\
75th Percentile\tablenotemark{f}, $z_{\rm s} \leq 1.0$ & 0.045 & 0.063 \\
Median Absolute Deviation\tablenotemark{f}, Excluding Catastrophic Outliers & 0.027 & 0.031\\
75th Percentile\tablenotemark{f}, Excluding Catastrophic Outliers & 0.045 &  0.054\\
\hline
\rule{0pt}{3ex}$|z_{\rm p} - z_{\rm c}| / (1 + z_{\rm c}) \leq 0.05$ and $m_{\rm F814W} \leq 22$ & 2109 / 4139 (51.0\%) & 2554 / 10684 (23.9\%) \\
\rule{0pt}{2ex}$|z_{\rm p} - z_{\rm c}| / (1 + z_{\rm c}) \leq 0.05$ and $m_{\rm F814W} \leq 24$ & 3612 / 10563 (34.3\%) & 4140 / 20263 (20.4\%)  \\
\rule{0pt}{2ex}$|z_{\rm p} - z_{\rm c}| / (1 + z_{\rm c}) \leq 0.05$ and $m_{\rm F814W} \leq 26$ & 4574 /20930 (21.9\%) & 6347 / 46338 (13.7\%) 
\enddata
\tablenotetext{a}{Photometric redshifts re-derived using the same techniques and parameters as in this work}
\tablenotetext{b}{$|(z_{\rm p} - z_{\rm s})| / (1 + z_{\rm s} ) \leq 0.05$}
\tablenotetext{c}{$0.15 \geq |(z_{\rm p} - z_{\rm s})| / (1 + z_{\rm s} ) > 0.05$}
\tablenotetext{d}{$0.5 \geq |(z_{\rm p} - z_{\rm s})| / (1 + z_{\rm s} ) > 0.15$}
\tablenotetext{e}{$|(z_{\rm p} - z_{\rm s})| / (1 + z_{\rm s} ) > 0.5$}
\tablenotetext{f}{$|(z_{\rm p} - z_{\rm s})| / (1 + z_{\rm s} )$}
\label{tab:photoz_comparison}
\end{deluxetable*}

After combining the spectral redshift catalogs and removing duplicates (any two objects with positions within 1$''$ of each other), we cross-match this catalog with our own catalog of detected objects. For each spectroscopic redshift, we match it with any object within $1\farcs5$ of the reported coordinates. In the event of multiple objects within this region, we match with the brightest object; if multiple objects are within 0.5 mag of this object, we match to the object closest to the spectroscopic position. From this matched catalog, we report 1716 objects with spectroscopic counterparts. For all galaxies, the median absolute redshift deviation between photometric and spectroscopic redshifts occurs at $\frac{|(z_{\rm p} - z_{\rm s})| }{ (1 + z_{\rm s} )} = 0.030$. In Table \ref{tab:photoz_comparison} we provide the median absolute deviations and 75th percentile deviations for the full galaxy sample, as well as the sample without catastrophic outliers and the sample with only galaxies with redshift $z_{\rm s} \leq 1.0$. The complete comparison of photometric and spectroscopic redshifts is shown in Figure \ref{fig:spec_to_phot_comp}. \\

\subsection{Rest-Frame Magnitudes}
\label{sect:SEDfit}

We fit SEDs to every galaxy in our catalog with ${\rm F814W} \leq 25.5$ mag (AB) or, for those galaxies outside of the F814W field of view, with apparent magnitude ${\rm m} \leq 25.5$ mag (AB) in the closest available filter. Stellar population models were fit to the data using \texttt{iSEDfit} \citep{2013ApJ...767...50M}, an IDL-based Bayesian inference routine for extracting physical parameters from broadband photometry. \texttt{iSEDfit} is a widely used code for estimating galaxy properties at all redshifts \citep[e.g.,][]{2012ApJ...747L...9Z, 2013ApJ...775...41A, 2013ApJ...779..138B, 2013ApJ...779..137Z, 2015ApJ...813..117F}. For all galaxies, we assume they are at the redshift of the cluster they are near.\\

To enable comparisons between clusters and with existing literature, we obtain rest-frame \textit{ugriz}, UBVRI \citep{1990PASP..102.1181B}, and JHK (from 2MASS filter curves) magnitudes for every object. We denote rest-frame magnitudes in this work with a leading superscript, e.g., ${}^{0.0}r$. To derive these values, \texttt{iSEDfit} begins by determining the closest matched filter to each target filter at the redshift of the galaxy. It then computes a magnitude offset based on the best-fit SED and returns the original photometry corrected by that offset. This procedure retains the photometric errors on the original filter. For galaxies lacking photometry in the best-matched filter, the code will synthesize a magnitude from the SED fit but will not return a photometric error, as those are limited to observed errors. Synthetic magnitudes are needed when galaxies are outside of the field of view of certain filters and therefore are not observed in the closest-matched filter. \\

\section{Comparison to Similar Works}

\subsection{Comparison to Previous CLASH Studies}
One way to identify potential systematic issues with our technique is to compare our results to other photometric studies of these clusters. However, the primary previous study using CLASH photometry \citep{2012ApJS..199...25P} was not tailored for optimizing cluster galaxy photometry; rather it was a general-purpose attempt to measure everything in the field, particularly background galaxies. Nevertheless, it provides an important first check of our results. \\

We match the publicly available photometric catalogs \citep{2012ApJS..199...25P} to spectroscopic redshifts following the same technique as we used to match our new results. To avoid biasing our results, we re-fit photometric redshifts to the public catalog using the same methods and parameters described in Section \ref{sect:phot_redshifts_described}. Due to the differing methodology in producing a detection catalog, there are slightly more galaxies with spectroscopic matches in the \citet{2012ApJS..199...25P} catalog than in ours. Using the same standards as in Section \ref{sect:phot_redshifts_described}, we compare the 1942 total matches to their spectroscopic counterparts in Table \ref{tab:photoz_comparison}. In addition to an almost $10\%$ drop in the percentage of well-matched galaxies, the old catalog also has ${\sim} 20\%$ to ${\sim} 40\%$ larger deviation between spectroscopic and photometric redshifts compared to our new work. The colors we present here provide a substantial increase in the accuracy of photometric redshifts for cluster members, although we note that \citet{2012ApJS..199...25P} presented accurate photometric redshifts for arcs, which are not in this analysis.\\

In comparison to the earlier catalog, our improved detection routine combined with more accurate photometric redshifts should produce a sample with greater purity for cluster identification. To test this claim, we consider the photometric redshifts of every object in these 25 fields, within three magnitude constraints. The fraction of galaxies with photometric redshifts close to the cluster redshift, such that $|(z_{\rm p} - z_{\rm c})| / (1 + z_{\rm c} ) \leq 0.05$, is significantly improved in our new catalog over the previous work; the exact fractions in three magnitude cuts are provided in Table \ref{tab:photoz_comparison}. As we do not need to segment objects to detect faint structure overlapping with extended halos, we produce a significantly less cluttered catalog.\\

Recent work by \citet{2017MNRAS.470...95M} produced another catalog of galaxies using the {\it HST} CLASH images, also demonstrating greatly improved photometric redshift accuracy over prior work by \citet{2014A&A...562A..86J}. Their improvement is a result of optimizing BPZ inputs. They also subtracted a parametric background model from the {\it HST} images to minimize the effects of ICL on photometry. However, the analysis of \citet{2017MNRAS.470...95M} is limited to the restrictive field of view of WFC3-IR, which only covers ${\sim} 40$\% of the area covered by CLASH ACS imaging. For cluster population studies, the known correlations between galaxy properties and local galaxy density (\citealt{1980ApJ...236..351D}; also see \citealt{1974ApJ...194....1O, 2004MNRAS.353..713K, 2010ApJ...721..193P}) or clustercentric radius \citep{1991ApJ...367...64W, 1993ApJ...407..489W, 2012ApJ...746..188M} imply that the significantly smaller fields of view of the  \citet{2017MNRAS.470...95M} catalogs limit their utility for any analysis of cluster galaxies. Additionally, that work uses Source Extractor for detection and photometry at only one scale, whereas in this work we perform a multi-scale, multi-color detection and compute galaxy-size-dependent local backgrounds/foregrounds.\\

\subsection{Comparison to the Hubble Frontier Fields}
Another comparison for our work is that of the Hubble Frontier Fields (HFF) catalogs produced by the ASTRODEEP collaboration \citep{2016A&A...590A..31C, 2016A&A...590A..30M}. They analyzed {\it HST} imaging of MACS-J0416 (as well as Abell 2744, which is outside the scope of the CLASH observations) using HFF {\it HST} data as well as ground-based K and {\it Spitzer} IRAC observations. HFF observations achieve a substantial depth, nominally ${\sim} 2$ mag fainter than CLASH data. The ASTRODEEP catalogs are created through a series of steps involving masking bright objects, fitting the ICL, and fitting bright galaxies, with intermediate steps involving subtraction of either ICL or galaxies. They make an excellent comparison sample for our work; their observations are deeper, their results have been refereed, and their catalog is compiled under different assumptions (galaxy and ICL model-based vs. model-agnostic).\\

To compare our results for MACS 0416 with those of ASTRODEEP, we first match our catalogs galaxy-by-galaxy. After sorting our catalog from brightest to faintest (using F814W magnitudes), we find a best match for each galaxy using the following process: if only one galaxy in the ASTRODEEP catalog is within $0\farcs65$ (10 pixels) of our target, that galaxy is matched with ours. If more than one possible match is within that angular radius, we consider only those galaxies with F814W magnitudes $\textrm{m} \leq \textrm{m}_\textrm{b} + 0.5$, where $\textrm{m}_\textrm{b}$ is the brightest galaxy in that angular range. We take the galaxy with smallest angular offset to the target in that subset to be the match.\\

\begin{figure*}
	\includegraphics[width=\textwidth]{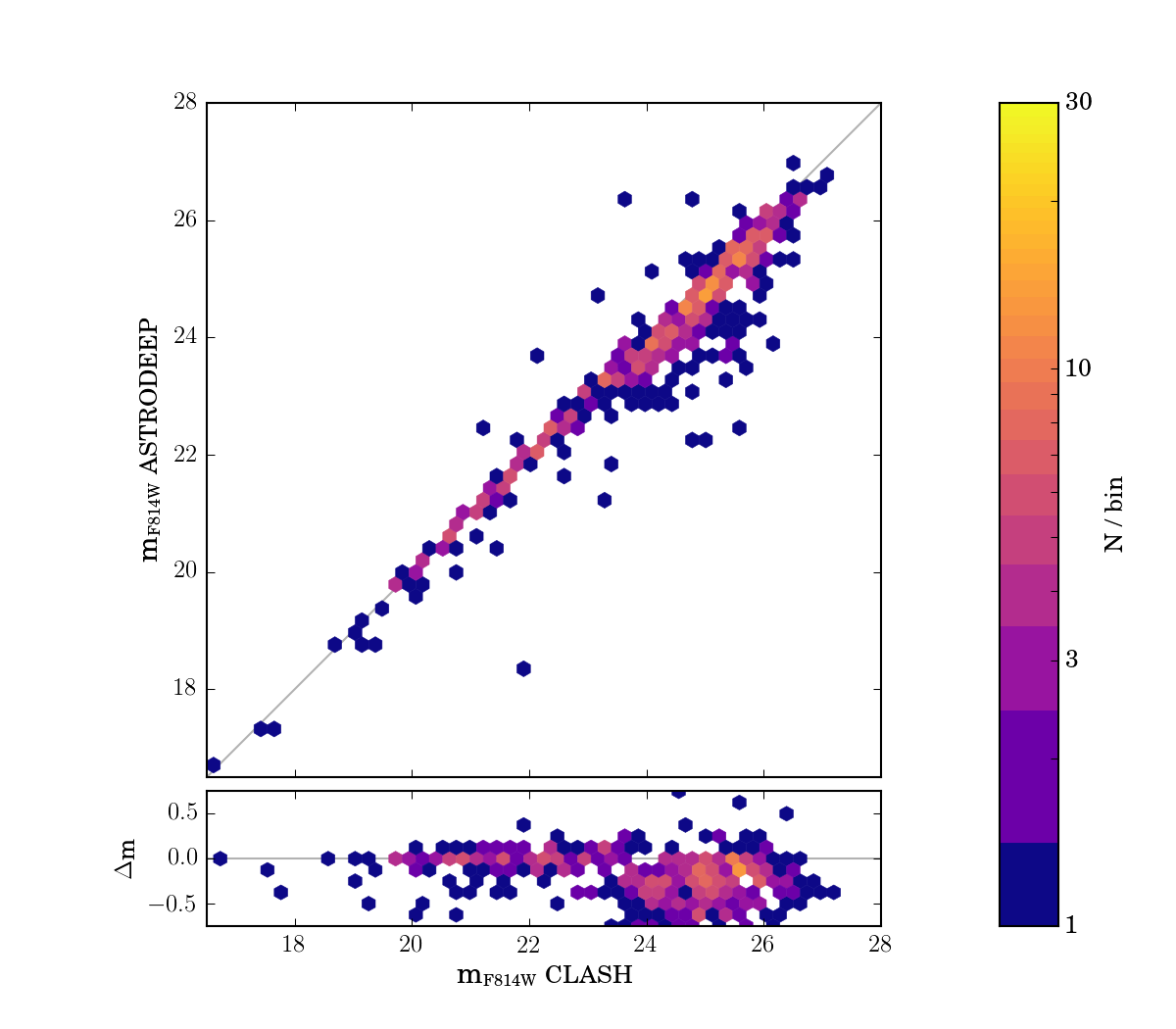}
    \caption{Comparison between F814W magnitudes measured by the ASTRODEEP Collaboration and this work for MACS 0416. Points are binned into hexagons, with the total density of points scaled logarithmically from 1 to 30 counts per hex. Details of the fit are provided in the text. A comparison of the offsets is provided in the lower panel.}
\label{fig:AD_F814W_comparison}
\end{figure*}

Our first cross-check between the two samples is to compare the measured magnitudes for each galaxy. This comparison is sensitive to both the ability of each method to capture the entire flux as well as how the background subtraction affects the faintest galaxies. This comparison is shown in Figure \ref{fig:AD_F814W_comparison}. Most galaxies have similar magnitudes reported in both studies, but we find slightly fainter magnitudes for the faintest galaxies; for magnitudes F814W $\geq 23$, the galaxies in this work are a median 0.23 mag fainter.\\

\begin{figure*}
\centering
	\includegraphics[width=0.8\textwidth]{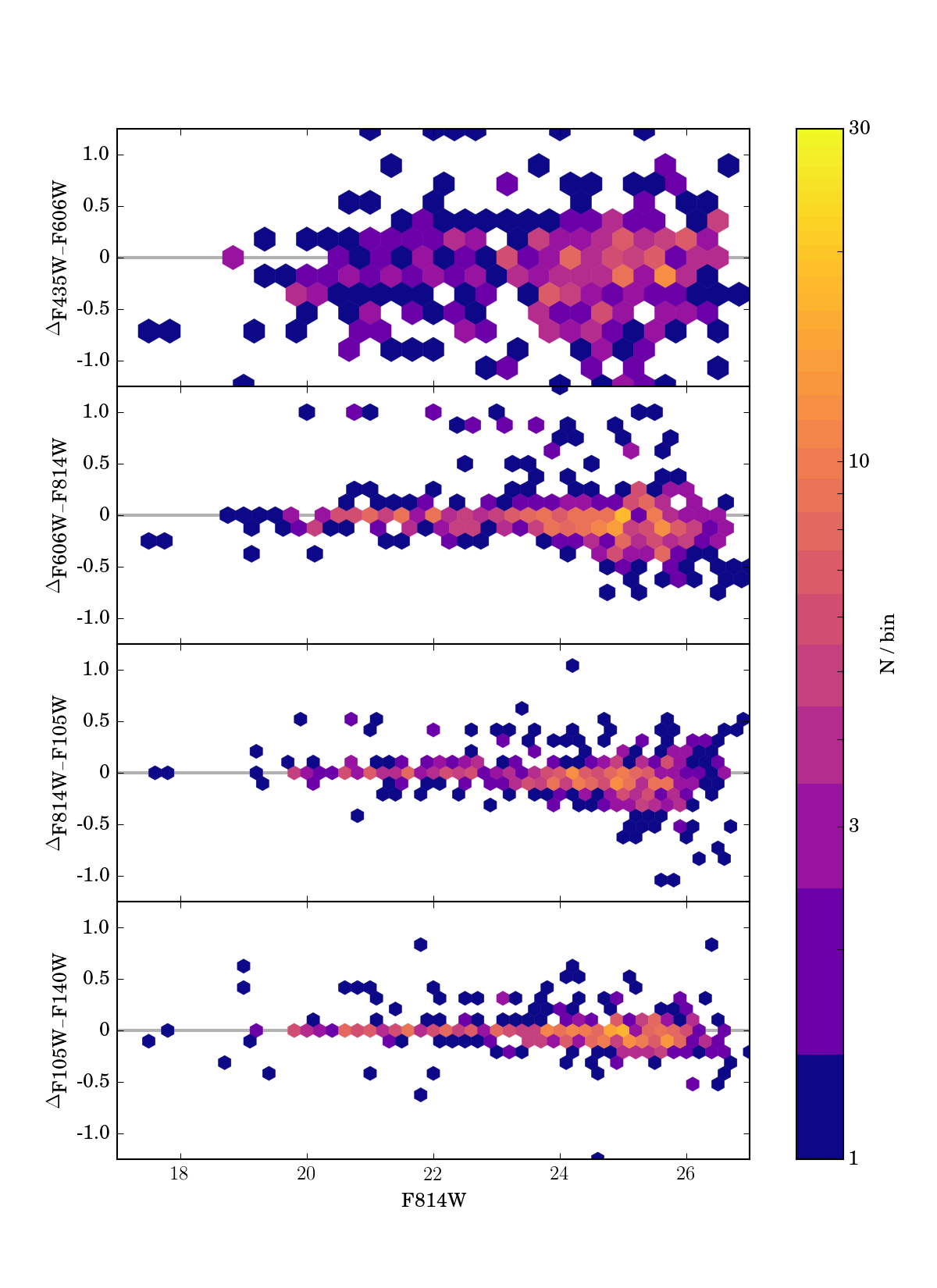}
    \caption{Comparison between measured colors for MACS 0416 in the ASTRODEEP catalog and this work. Counts are binned into hexagons, with the total density of points scaled logarithmically. Note that the hex size is increased for bluer colors.}
\label{fig:AD_color_comparison}
\end{figure*}

We next consider how galaxy colors differ between the two techniques. While the HFF observations are deeper than those of CLASH, they only have seven {\it HST} filters, so the color comparison is limited. Figure \ref{fig:AD_color_comparison} shows the difference between measured colors between this work and ASTRODEEP. We see a slight trend for galaxies in our sample to be bluer in F435W-F606W than in the ASTRODEEP catalog, but that color discrepancy is diminished for all of the redder filters. Numerically, the median offset (defined such that a negative offset corresponds to a bluer object in our catalog) between colors is
\begin{enumerate}
\item $\langle \Delta_{({\rm F435W} - {\rm F606W})}\rangle = -0.137 \pm 0.396$
\item $\langle \Delta_{({\rm F606W} - {\rm F814W})}\rangle = -0.060 \pm 0.129$
\item $\langle \Delta_{({\rm F814W} - {\rm F105W})}\rangle = -0.037 \pm 0.112$
\item $\langle \Delta_{({\rm F105W} - {\rm F140W})}\rangle = -0.026 \pm 0.068$
\end{enumerate}
Here the reported uncertainties are 1.4286$\times$ the median absolute deviation and all values are in magnitudes. For galaxies with photometric redshift offset from the cluster $\vert z_p - z_c \vert $ \textless 0.1 and red filter magnitude \textless \ 25 mag (AB), we report the same set of values, as well as the median color uncertainty for our photometry alone:
\begin{enumerate}
\item $\langle \Delta_{({\rm F435W} - {\rm F606W})}\rangle = -0.162 \pm 0.442$, $\langle \sigma \rangle = 0.171$
\item $\langle \Delta_{({\rm F606W} - {\rm F814W})}\rangle = -0.036 \pm 0.079$, $\langle \sigma \rangle = 0.033  $
\item $\langle \Delta_{({\rm F814W} - {\rm F105W})}\rangle = +0.004 \pm 0.057$, $\langle \sigma \rangle = 0.018  $
\item $\langle \Delta_{({\rm F105W} - {\rm F140W})}\rangle = -0.012 \pm 0.026$, $\langle \sigma \rangle = 0.013 $
\end{enumerate}
Again, all values are in magnitudes. \\

There is no significant offset in colors between these two works, although the deviation is larger than expected from measurement errors alone. One possible source of the scatter is from different galaxy apertures, which would influence the colors of galaxies that have a color gradient. While ASTRODEEP did not publish their detection areas, we investigate the effects of size differences by limiting our comparison to galaxies with measured brightnesses within 0.1 mag of each other in one filter. While we were able to reduce measured deviations to the order of the calculated uncertainties for the bluest two colors through magnitude constraints, there remained a larger spread in color offsets for the redder colors, particularly among cluster galaxies, than could be explained by color uncertainties alone. The color differences appear to therefore be real, and caused by the different methodologies for calculating a photometric background. Nevertheless, as the largest color differences involve the F435W and F606W filters, we do not expect the differing colors to cause significant variations in SED fitting between the two reductions -- at $z = 0.397$ (the redshift of MACS 0416), these filters cover the range below the $4,000\ {\rm \AA}$\ break for cluster galaxies.\\

\begin{figure*}
		\includegraphics[width=\textwidth]{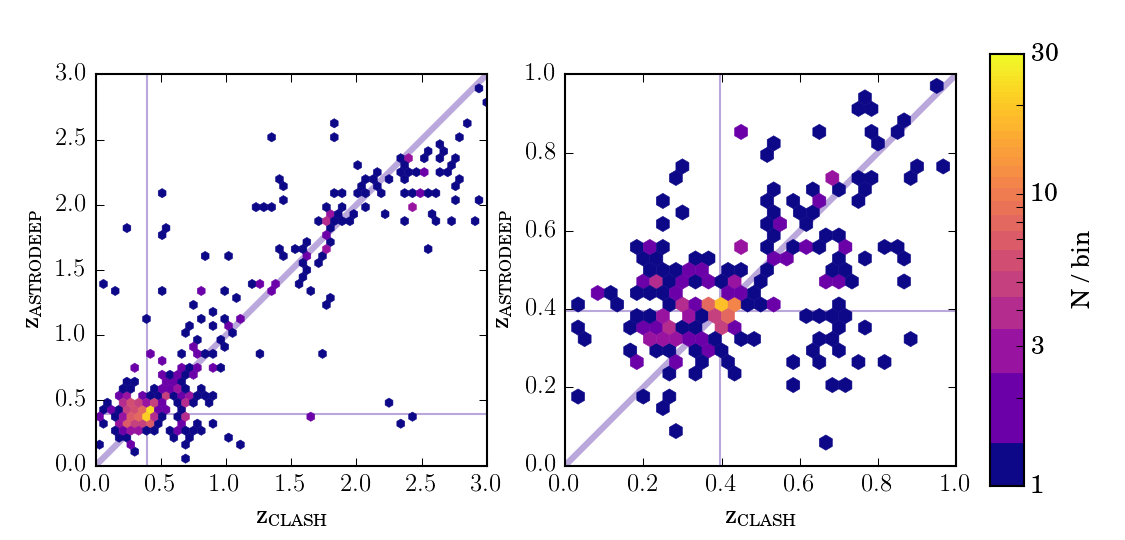}
    \caption{Comparison between redshifts measured by the ASTRODEEP Collaboration and this work for MACS 0416. Points are binned into hexagons, with the total density of points scaled logarithmically from 1 to 30 counts per hex, as indicated by the colorbar. Details of the fit are provided in the text. A zoom-in to just matches below redshift 1.0 is provided in the right panel. The cluster redshift is indicated by the vertical and horizontal lines; the diagonal line is the identity line.}
\label{fig:AD_Z_comparison}
\end{figure*}

One way to quantify the importance of these color offsets is to cross-check the photometric redshifts reported here and by ASTRODEEP. The latter used multiple methods for redshift estimation, including using spectroscopic redshifts where available, and combined them all for a final answer, while we only consider our redshift estimates from BPZ. We show in Figure \ref{fig:AD_Z_comparison} a comparison between our reported photometric redshifts and the photometric and spectroscopic redshifts reported by the ASTRODEEP collaboration. When considering the uncertainties on these measurements, we find good agreement between our reported redshifts and the comparison sample. For galaxies with photometric redshifts $z_p < 1.0$ in either of our samples and brighter than 25th magnitude in the F814W filter, the median redshift offset between our catalogs is 0.004, with an uncertainty ($1.4826 \times$ the median absolute deviation) of $\pm$ 0.188.\\

We have compared the results of our mode-measuring background calculation technique to the results of a more traditional background-modeling photometry routine for one cluster. Despite comparing CLASH observations to deeper HFF observations, we see no significant offset in photometry. While further work is still needed to understand the systematic issues inherent to these and other photometric techniques, the comparisons presented here suggest that there is no significant offset between parametric structure modeling and local modal background estimation.\\

\section{Cluster Membership}
\label{sect:Members}

Our next step is to define membership criteria for selection of cluster members. Determining cluster membership is necessary for, among other things, measuring cluster luminosity functions. We have already calculated a redshift probability distribution and an SED goodness-of-fit at the cluster redshift for all of our galaxies, and we supplement these calculations with measured spectroscopic redshifts where available. We describe how we use this information to select cluster members below; we then compare how the sample completeness of cluster galaxies differs between our criteria and through cluster membership selection of only galaxies along the red sequence.\\

For the sake of homogeneity and uniform data quality, we investigate membership, and the properties of candidate cluster members, down to ${\rm F814W} \leq 25.5$ mag (AB). We consider the photometry, photometric redshift parameters, SED fit values, rest-frame magnitudes, and spectroscopic redshift measurements for all of these galaxies in our membership selection. For each galaxy, we use the discrete probability distributions produced by BPZ to determine a total probability of that galaxy being within some redshift range of the nominal cluster redshift. Here, we consider $|\Delta_{\rm z}| < 0.03$, $|\Delta_{\rm z}| < 0.05$, $|\Delta_{\rm z}|/(1 + {\rm z}_{\rm c}) < 0.03$, and $|\Delta_{\rm z}|/(1 + {\rm z}_{\rm c}) < 0.05$; we label the summed probabilities within those ranges ${\rm P}_{03}$, ${\rm P}_{05}$, ${\rm P}_{103}$, and ${\rm P}_{105}$, respectively.\\

To determine membership, we step each galaxy through a series of screens; the result of each step is that a galaxy is classified as a member, classified as a non-member, or passed on to the next step. The first sieve is to select cluster galaxies by their spectroscopic redshifts. Any galaxy with $|\Delta_{\rm z}|/(1 + {\rm z}_{\rm c}) < 0.03$ is considered a cluster member; those with $|\Delta_{\rm z}|/(1 + {\rm z}_{\rm c}) > 0.10$ are considered non-members. The rest of the galaxies -- either those with indeterminate or no spectroscopic redshifts -- are then characterized by their photometric redshift probabilities. As a first pass, those galaxies with total probability ${\rm P}_{03} > 0.8$ are assigned as members, while those with ${\rm P}_{105} < 0.1$ are classified as non-members. \\

We next screen for cluster members by selecting those with a well-fit SED or a best-fit photometric redshift solution consistent with being in a cluster. Galaxies with $\chi^2_\nu < 1.5$ and $\chi^2_\nu > 0.7$ (to avoid selecting galaxies with poorly constrained fits through this cut) in the SED fit at the cluster redshift are classified as cluster members, as are those with a most likely or best redshift determination from BPZ within $|\Delta_{\rm z}|/(1 + {\rm z}_{\rm c}) < 0.05$. For the remaining objects, we examine the distributions of ${\rm P}_{03}$, ${\rm P}_{05}$, ${\rm P}_{103}$, and ${\rm P}_{105}$; we only classify those remaining galaxies with ${\rm P}_{103} > 0.2$ and ${\rm P}_{105} > 0.6$ as members, while the rest are classified as non-members. \\

An alternative way to select cluster members is to select only those galaxies with colors that align with the identified red sequence in a cluster \citep[e.g.,][among many others]{2000AJ....120.2148G, 2007ApJ...660..221K, 2009ApJ...702..745H, 2014ApJ...785..104R, 2016ApJS..224....1R}. Red sequence selection is a powerful technique for many applications but is problematic for others. For example, the color selection criteria can be overly restrictive and exclude galaxies with recent star formation, such as, e.g., Butcher--Oemler \citep{1978ApJ...219...18B} or  Dressler--Gunn \citep{1983ApJ...270....7D} galaxies. Therefore, we quantify the sample completeness as functions of magnitude and color below. \\

\begin{figure*}
	\includegraphics[width=\textwidth]{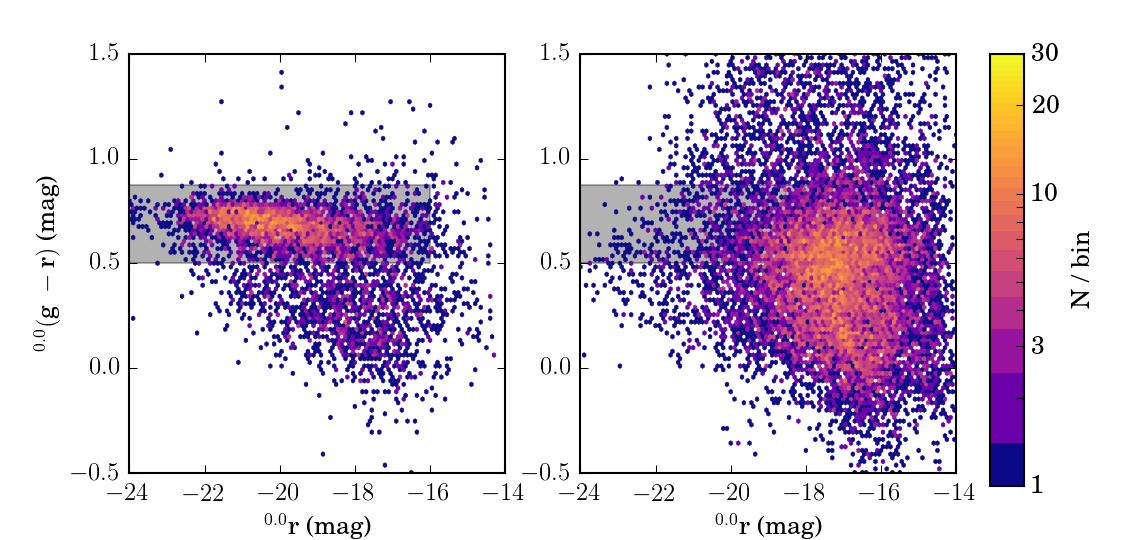}
    \caption{Rest-frame $^{0.0}(g-r)$ CMD for all 25 clusters. Shown on the left are those galaxies we call members, as presented in Section \ref{sect:Members}, while those classified as non-members are plotted on the right. A potential red sequence selection region is shown as a shaded box. Hex bins are scaled logarithmically with the number of galaxies contained inside.}
\label{fig:membership_missing_ccd}
\end{figure*}

In Figure \ref{fig:membership_missing_ccd} we show a plot of  $^{0.0}(g-r)$ colors using rest-frame magnitudes, as described in Section \ref{sect:SEDfit}. Here, we define a selection region, counting those galaxies with $0.5 < {}^{0.0}(g-r) < 0.875$. For galaxies brighter than $^{0.0}r < -16$, 70.4\% of cluster members are inside this color region, but so is an additional population of non-members with size equal to 75.2\% of the total member population above that brightness threshold. Increasing the magnitude cut to $^{0.0}r < -20$, 89.1\% of members fall within that color region, while the contaminant population is only equal to 20.0\% of the total cluster population in that luminosity range. Based on these results, selecting galaxies using the red sequence is well suited for selecting galaxies at the tip of the relation. However, this selection not only fails to account for the entire cluster population at fainter magnitudes, it also becomes significantly affected by contamination.\\

\section{Luminosity Function}

Having assembled a catalog of cluster galaxies with magnitudes adjusted to the same band via SED-fitting, we investigate the variations in cluster populations by comparing their luminosity functions. Here, we only consider the luminosity functions of the entire cluster populations (we do not fit luminosity functions to just the red or blue populations) in only one band. Likewise, we do not correct for the different metric fields of view for each cluster. We will present a more comprehensive analysis of the CLASH luminosity functions in a future work.\\

\begin{figure*}
	\includegraphics[width=\textwidth]{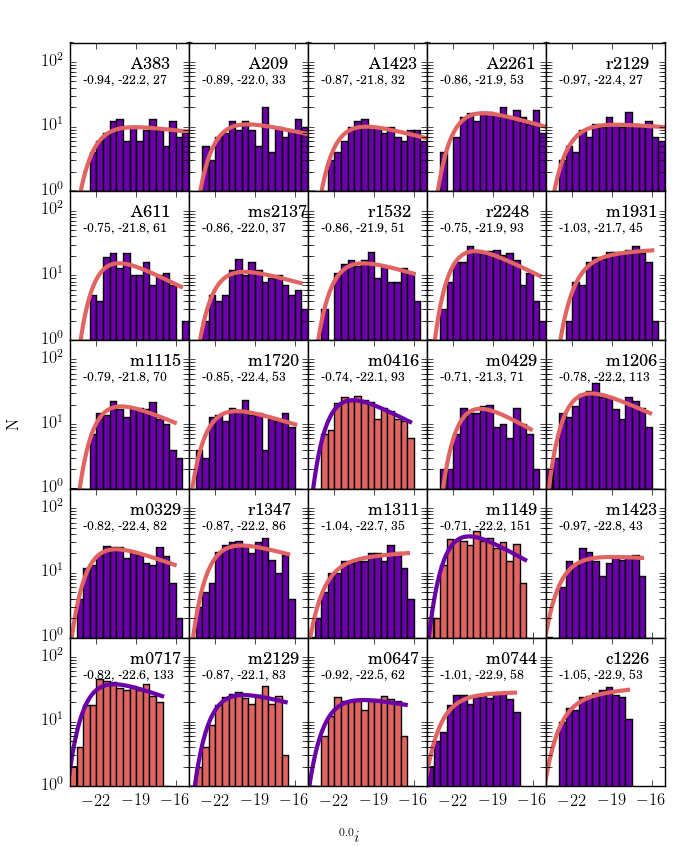}
    \caption{Rest-frame $^{0.0}i$-magnitude luminosity functions for all 25 CLASH clusters. Clusters are shown binned in half-magnitude intervals. The best-fit Schechter luminosity function is shown in orange (purple) for X-ray-selected (high-magnification) clusters, while the binned galaxy counts are shown in purple (orange). Clusters are plotted in order of increasing redshift. Best-fit values of $\alpha$, $M^*$, and $\phi^*$ are provided in that order on each plot. Details of the fits are provided in the text.}
\label{fig:all_lumin_functions}
\end{figure*}

\begin{deluxetable*}{lrrrr}
\tabletypesize{0.6\scriptsize}
\tablecaption{CLASH Luminosity Function Fit}
\tablewidth{0pt}
\tablehead{
\colhead{Cluster Name} & \colhead{$\alpha$} & \colhead{${\rm M}_i^*$} & \colhead{$\phi^*$} & \colhead{Det. Limit} \\
\colhead{} & \colhead{} & \colhead{(mag)} & \colhead{} &  \colhead{(mag)}
}
\startdata
Abell 383  \rule{0pt}{3ex}  & $-0.94^{+ 0.05}_{- 0.05}$  & $-22.17^{+  0.26}_{-  0.35}$  & $ 26.6^{+  6.2}_{-  5.9}$  & $M^* +  7.32$ \\
Abell 209  \rule{0pt}{3ex}  & $-0.89^{+ 0.05}_{- 0.05}$  & $-22.02^{+  0.24}_{-  0.27}$  & $ 33.2^{+  7.1}_{-  6.6}$  & $M^* +  6.97$ \\
Abell 1423  \rule{0pt}{3ex}  & $-0.87^{+ 0.06}_{- 0.05}$  & $-21.85^{+  0.25}_{-  0.26}$  & $ 31.9^{+  7.6}_{-  5.9}$  & $M^* +  7.38$ \\
Abell 2261  \rule{0pt}{3ex}  & $-0.86^{+ 0.04}_{- 0.04}$  & $-21.86^{+  0.20}_{-  0.21}$  & $ 53.3^{+  9.2}_{-  8.4}$  & $M^* +  6.83$ \\
RXJ 2129  \rule{0pt}{3ex}  & $-0.97^{+ 0.05}_{- 0.05}$  & $-22.35^{+  0.27}_{-  0.34}$  & $ 27.0^{+  6.9}_{-  5.7}$  & $M^* +  6.80$ \\
Abell 611  \rule{0pt}{3ex}  & $-0.75^{+ 0.08}_{- 0.05}$  & $-21.79^{+  0.22}_{-  0.18}$  & $ 60.8^{+ 14.6}_{-  8.5}$  & $M^* +  5.58$ \\
MS 2137  \rule{0pt}{3ex}  & $-0.86^{+ 0.06}_{- 0.07}$  & $-21.97^{+  0.22}_{-  0.33}$  & $ 37.0^{+  8.3}_{-  8.5}$  & $M^* +  6.50$ \\
RXJ 1532  \rule{0pt}{3ex}  & $-0.86^{+ 0.06}_{- 0.07}$  & $-21.92^{+  0.20}_{-  0.22}$  & $ 51.0^{+ 10.6}_{-  9.5}$  & $M^* +  5.85$ \\
RXJ 2248  \rule{0pt}{3ex}  & $-0.75^{+ 0.04}_{- 0.06}$  & $-21.87^{+  0.14}_{-  0.21}$  & $ 93.3^{+ 11.7}_{- 15.5}$  & $M^* +  5.76$ \\
MACS 1931  \rule{0pt}{3ex}  & $-1.03^{+ 0.06}_{- 0.05}$  & $-21.72^{+  0.21}_{-  0.24}$  & $ 45.3^{+ 11.4}_{-  7.9}$  & $M^* +  5.06$ \\
MACS 1115  \rule{0pt}{3ex}  & $-0.79^{+ 0.05}_{- 0.08}$  & $-21.83^{+  0.17}_{-  0.23}$  & $ 69.7^{+ 10.8}_{- 12.3}$  & $M^* +  4.88$ \\
MACS 1720  \rule{0pt}{3ex}  & $-0.85^{+ 0.07}_{- 0.05}$  & $-22.43^{+  0.21}_{-  0.22}$  & $ 53.5^{+ 12.2}_{-  8.4}$  & $M^* +  6.02$ \\
MACS 0416  \rule{0pt}{3ex}  & $-0.74^{+ 0.05}_{- 0.07}$  & $-22.07^{+  0.15}_{-  0.21}$  & $ 93.5^{+ 13.0}_{- 15.3}$  & $M^* +  5.71$ \\
MACS 0429  \rule{0pt}{3ex}  & $-0.71^{+ 0.06}_{- 0.10}$  & $-21.32^{+  0.16}_{-  0.31}$  & $ 71.5^{+ 10.7}_{- 16.0}$  & $M^* +  5.08$ \\
MACS 1206  \rule{0pt}{3ex}  & $-0.78^{+ 0.05}_{- 0.05}$  & $-22.21^{+  0.14}_{-  0.16}$  & $113.2^{+ 15.0}_{- 16.0}$  & $M^* +  5.83$ \\
MACS 0329  \rule{0pt}{3ex}  & $-0.82^{+ 0.05}_{- 0.06}$  & $-22.40^{+  0.17}_{-  0.19}$  & $ 81.7^{+ 11.8}_{- 14.0}$  & $M^* +  5.98$ \\
RXJ 1347  \rule{0pt}{3ex}  & $-0.87^{+ 0.07}_{- 0.04}$  & $-22.22^{+  0.17}_{-  0.17}$  & $ 86.0^{+ 17.2}_{- 11.2}$  & $M^* +  5.50$ \\
MACS 1311  \rule{0pt}{3ex}  & $-1.04^{+ 0.05}_{- 0.06}$  & $-22.73^{+  0.24}_{-  0.41}$  & $ 35.4^{+  7.9}_{-  8.4}$  & $M^* +  5.88$ \\
MACS 1149  \rule{0pt}{3ex}  & $-0.71^{+ 0.04}_{- 0.06}$  & $-22.18^{+  0.12}_{-  0.18}$  & $151.3^{+ 15.6}_{- 21.9}$  & $M^* +  5.34$ \\
MACS 1423  \rule{0pt}{3ex}  & $-0.97^{+ 0.06}_{- 0.06}$  & $-22.85^{+  0.24}_{-  0.39}$  & $ 43.5^{+ 10.0}_{-  9.4}$  & $M^* +  5.83$ \\
MACS 0717  \rule{0pt}{3ex}  & $-0.82^{+ 0.05}_{- 0.05}$  & $-22.56^{+  0.14}_{-  0.16}$  & $133.1^{+ 17.6}_{- 16.8}$  & $M^* +  5.04$ \\
MACS 2129  \rule{0pt}{3ex}  & $-0.87^{+ 0.06}_{- 0.06}$  & $-22.11^{+  0.16}_{-  0.18}$  & $ 82.5^{+ 14.7}_{- 14.0}$  & $M^* +  4.95$ \\
MACS 0647  \rule{0pt}{3ex}  & $-0.92^{+ 0.07}_{- 0.05}$  & $-22.55^{+  0.20}_{-  0.21}$  & $ 62.5^{+ 14.1}_{-  9.4}$  & $M^* +  5.58$ \\
MACS 0744  \rule{0pt}{3ex}  & $-1.01^{+ 0.06}_{- 0.06}$  & $-22.88^{+  0.21}_{-  0.28}$  & $ 57.7^{+ 12.5}_{- 12.9}$  & $M^* +  5.10$ \\
CLJ 1226  \rule{0pt}{3ex}  & $-1.05^{+ 0.07}_{- 0.06}$  & $-22.86^{+  0.33}_{-  0.25}$  & $ 53.1^{+ 13.4}_{- 11.2}$  & $M^* +  4.52$ \\
\hline\\
$0.0 < z < 0.32  \rule{0pt}{3ex} $  & $-0.90^{+ 0.02}_{- 0.04}$  & $-22.04^{+  0.09}_{-  0.13}$  & $ 35.8^{+  2.5}_{-  4.2}$   & -16.00 \\
$0.32 < z < 0.4  \rule{0pt}{3ex} $  & $-0.87^{+ 0.04}_{- 0.02}$  & $-22.05^{+  0.10}_{-  0.07}$  & $ 58.9^{+  6.8}_{-  3.0}$   & -16.50 \\
$0.4 < z < 0.55  \rule{0pt}{3ex} $  & $-0.84^{+ 0.01}_{- 0.04}$  & $-22.41^{+  0.05}_{-  0.10}$  & $ 91.0^{+  3.6}_{-  9.3}$   & -17.00 \\
\enddata 
\label{tab:CLASH_luminosity_function}
\end{deluxetable*}

We fit our data with a Schechter luminosity function \citep{1976ApJ...203..297S}, of the form
\begin{equation}
\begin{split}
\phi(M)\ dM = &0.4\ \phi^* \ln(10) \times 10^{0.4 (M^* - M) \times (1 + \alpha)}\\
& \times \exp( -10^{0.4 (M^* - M)})\ dM,
\end{split}
\end{equation}
where $\phi^*$ is the number of galaxies per magnitude, $M^*$ is the characteristic magnitude of the function, $M$ is the absolute magnitude of a galaxy, and $\alpha$ is the faint-end slope. We remind the reader that, in this form, a ``flat slope'' occurs at $\alpha = -1$. As we have made no attempt to adjust for the volumes sampled by each cluster, $\phi^*$ is a normalization and not a density. For each cluster, we determine the parameters of the Schechter function using unbinned luminosity data, by following the maximum likelihood technique described in \cite{1983ApJ...269...35M} and \cite{1999ApJ...523L.137D}. Using interpolated rest-frame $^{0.0}i$-band magnitudes, we find the minimum of a likelihood function defined as
\begin{equation}
\label{eqn:gof_schechter}
\begin{split}
S &= -2 \ln (\mathcal{L}) \\
&= -2 \sum\limits_{\rm i}^{\rm N} \ln(\ \phi(\ {\rm M}_{\rm i})) + 2 \int\limits_{ {\rm M}_{\rm faint}} \phi({\rm M})\ dM
\end{split}
\end{equation}
in a grid of values for $\alpha$, $M^*$ and $\phi^*$. While $S$ itself does not provide a goodness-of-fit statistic in the same way as $\chi^2$, the errors in individual parameters are still bound by contours in $\Delta S$ in the same way as they would be for $\Delta \chi^2$.\\

${\rm M}_{\rm faint}$ is set for each cluster to be the faintest object classified as a member. We use a 50x50x25 grid uniformly covering $\alpha {=} [-1.3, -0.5]$,  $M^* {=} [-23, -18]$, and $\phi^* {=} [10^{-1}, 10^{4}]$, where $\phi^*$ is in units of galaxies per magnitude per cluster field (within the observed field of view). After finding the best value, we examine a sub-grid covering the 9x9x9 box centered on the best parameters. We repeat the sub-sampling step a second time, for a total of three resolution levels for the three luminosity function parameters of interest. To determine the $1 \sigma$ uncertainties we find the largest and smallest value of each parameter in the $\alpha$, $M^*$, $\phi^*$ parameter space in which $\Delta(S) \leq 1.00$. \\ 

We present our best-fit values for each of the CLASH clusters in Table \ref{tab:CLASH_luminosity_function}. Uncertainties listed are $1 \sigma$ values. We plot our luminosity functions, as well as histograms of cluster members, in Figure \ref{fig:all_lumin_functions}. $\phi^*$ has been divided by 2 in this figure to match the half-magnitude-wide bins. We also fit the combined galaxy sample in three redshift bins, each containing seven clusters. The results of those fits, as well as the values of ${\rm M}_{\rm faint}$ used in the fits, are shown in Table \ref{tab:CLASH_luminosity_function}. For the combined samples, the values of $\phi^*$ presented in Table \ref{tab:CLASH_luminosity_function} have been divided by 7. \\

\begin{figure*}
	\includegraphics[width=\textwidth]{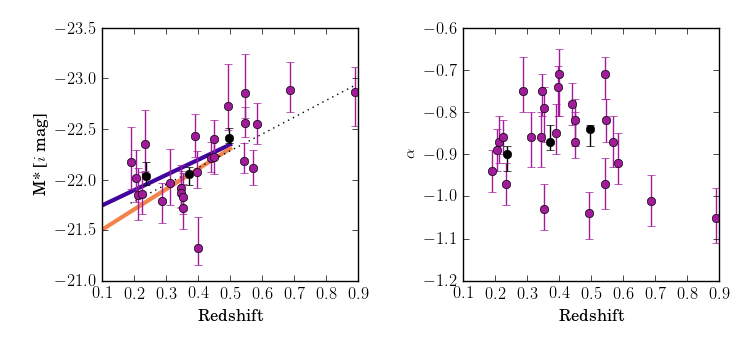}
    \caption{Left: measured values of $M^*$ for {\it i}-band luminosity functions of CLASH clusters fit individually (purple) and in redshift bins (black). The dotted black line traces the best fit of the redshift evolution of $M^*$, as described in the text. Shown in blue and orange are fits for $i$-band $M^*$ values for all (blue) and only red (orange) galaxies, with evolution parameters from \cite{2012MNRAS.420.1239L}. Right: measured values of $\alpha$ for $i$-band luminosity functions of CLASH clusters fit individually (purple) and in redshift bins (black).}
\label{fig:all_Mstar_values}
\end{figure*}

In Figure \ref{fig:all_Mstar_values} we plot our measured values of $M^*$ in $i$-band for all clusters (purple) as a function of redshift, as well as for the three binned fits to the luminosity function, set to the average redshift of the binned clusters (black). There is a trend for increasing $M^*$ brightness with increasing redshift, which is not unexpected \citep[e.g.][]{2006ApJ...652..249X}. We perform a linear regression of our measured values of $M^*$ to a line of the form
\begin{equation}
 M^*(z) = a \times z + M^*_0.
 \end{equation}
The best-fit values of this regression are $a = -1.65$ and $M^*_0 = -21.45$. This fit is shown in Figure \ref{fig:all_Mstar_values} by the dotted black line. We also plot the best-fit evolutionary parameters from the Galaxy and Mass Assembly (GAMA) k-corrected $i$-band luminosity fuctions presented by \cite{2012MNRAS.420.1239L}, after adjusting their reported values of $M^*$ to an $H_0 = 70\ {\rm km} \ {\rm s}^{-1}$ cosmology. Figure \ref{fig:all_Mstar_values} shows the evolution of $M^*$ both for all galaxies (blue line) and only red galaxies (orange line) reported by \cite{2012MNRAS.420.1239L}. \\

When integrating the Schechter function to find the likelihood as described by Equation \ref{eqn:gof_schechter}, we set ${\rm M}_{\rm faint}$ equal to the faintest cluster member. However, we were also interested in the limit at which the Schechter function no longer described the observed galaxy counts in terms of ${\rm M} - {\rm M}^* $. To determine this limit, we found the faintest half-magnitude-wide bin in which the number of galaxies was within 90\% of that expected by the cluster's luminosity function. These values are reported for each cluster in Table \ref{tab:CLASH_luminosity_function}. We find that the luminosity functions hold for ${\sim} 5$ mag fainter than $M^*$ for the entire sample and to ${\sim} M^* + 7$ for the lowest-redshift clusters. Previous works that have probed the luminosity function of clusters to this depth have been mostly limited to low redshift \citep[e.g.,][]{2005A&A...434..521S, 2006AJ....132..347C, 2009MNRAS.396.2367S, 2012AJ....144...40Y}, with the notable exception of  \cite{2013MNRAS.434.3469D}, who obtained a similar depth in three combined samples containing fewer total clusters than studied here.\\

For our sample, we find best-fit $\alpha$ values in the range $-1.0 \lesssim \alpha \lesssim -0.7$ and $M^*$ values in the range $-23 \lesssim {\rm M}^* \lesssim -21.5$; when considering the binned samples, we find that $\alpha$ declines from -0.90 to -0.84 and $M^*$ brightens from -22.0 to -22.4 in the redshift range sampled here. We compare these values to those of other works, beginning with \cite{2009ApJ...700.1559R}. They fit SDSS clusters drawn from the sample presented in \cite{2007MNRAS.379..867V} to a maximum redshift of $z \leqslant 0.06$. For $i$-band observations, their best fit was ${\rm M}^{*}_{i} = -21.46^{+0.03}_{-0.04}$, $\alpha = -0.75^{+0.02}_{-0.01}$. That work also  examined the luminosity function of 16 clusters in the ESO Distant Cluster Survey (EDisCS) spanning redshifts $0.4 < z < 0.8$. When only including red sequence members, the EDisCS clusters together were best fit by an {\it i}-band luminosity function of ${\rm M}^{*}_{i} = -21.80^{+0.22}_{-0.17}$ and $\alpha = -0.34^{+0.16}_{-0.10}$. \\

\cite{2013MNRAS.434.3469D} studied 11 merging clusters at $0.2 \lesssim z \lesssim 0.6$, finding best-fit Schechter slopes of $\alpha \approx -1$. \cite{2007ApJ...661...95S} derived values of $\alpha = -0.91 \pm 0.02$, ${\rm M}_{\rm V} = -21.39 \pm 0.05$ for 10 MACS \citep{2001ApJ...553..668E} clusters at $z \sim 0.5$, including several considered in this work. \cite{2015A&A...575A.116M} presented measurements of luminosity functions for $0.4 \leq z < 0.9$ at rest-frame I and R; their red-sequence-selected results are $\alpha_{\rm R} = -0.80 \pm 0.14$, ${\rm M}^*_{\rm R} = -22.4 \pm 0.2$ and $\alpha_{\rm I} = -0.37 \pm 0.18$, ${\rm M}^*_{\rm I} = -22.0 \pm 0.2$. \citet{2017A&A...604A..80M} looked at a subset of CLASH clusters using only two filters of {\it HST} imaging and selecting all galaxies along the red sequence to a shallower depth than explored here; in two redshift bins, they find a steepening slope ($\alpha = -0.85 \pm 0.13$ at $z = 0.289$ to $\alpha = -0.63 \pm 0.17$ at $z = 0.512$) for {\it HST} data. Using ground-based imaging of two CLASH clusters to measure the stellar mass function of all galaxies beyond $r_{200}$, \citet{2014A&A...571A..80A, 2016A&A...585A.160A} found slopes of $\alpha = -0.85 \pm 0.04$ and $\alpha = -1.17 \pm 0.02$.\\

The results in these works are consistent with what is presented here. We see a passive evolution in $M^*$ with redshift and no statistically significant change in $\alpha$ with redshift. Several works report a much steeper slope (lower value of $\alpha$), namely \citet{2017A&A...604A..80M}, \cite{2015A&A...575A.116M}, and \cite{2009ApJ...700.1559R}, but these fits only considered red sequence galaxies. While several other studies have found steeper faint-end slopes for red sequence-only samples (e.g., \citealt{2004ApJ...610L..77D}; but see also \citealt{2006MNRAS.369..969A, 2008MNRAS.386.1045A, 2014A&A...565A.120A}) works that do not select only red sequence members find $\alpha$ values more consistent with what we present here \citep{2010A&A...524A..17S, 2012ApJ...761..141M}. \\

\section{Summary}

We have presented a new framework for detection and photometry of galaxies in crowded fields. Using the computational power of a desktop computer and the maps of observed variance in pixel fluxes from {\it HST} observations of the CLASH fields, we characterize the background local to each galaxy on a scale appropriate to that galaxy's size. As that local background may include the flux of a nearby galaxy, we work galaxy-by-galaxy, leaving behind the measured background as we go. This allows us to accurately photometer overlapping galaxies.\\

We also present a technique for detecting galaxies in a cluster by mapping structure at all scales in an image. By connecting the large-scale structure in an image with the peaks of flux distributions, we can detect galaxies at their full extent; however, by looking for small-scale structure, we can identify smaller galaxies that are otherwise buried in the wings of larger, brighter galaxies. This technique results in a significantly improved fraction of cluster members in the sample of detected galaxies, as small galaxies can be recovered without needing to fragment large galaxies. We provide one caution for using the same methods to detect sources in other data. As the same statistical invariance across large scales that causes light from BCGs and the ICL to be removed around other galaxies also exists in resolved spiral galaxies, detection catalogs that successfully identify small cluster members will also detect individual knots of stars in late-type galaxies. Without filtering, a detection catalog will over-count foreground galaxy populations and also not correctly photometer the resolved foreground spiral galaxies that actually do exist.\\

As future observatories come online, and as new deep surveys of cluster environments are undertaken, it will be necessary for photometric pipelines to account for the issues of crowded environments in order to maximize the return on investment for these programs. With the CLASH images, we have demonstrated the ability of mode-based background determination to detect and photometer galaxies of all sizes across deep, multi-wavelength datasets. We briefly summarize the catalog of CLASH galaxies we have produced and the initial scientific results of that survey.

\begin{enumerate}
\item We detect and photometer 20,930 objects in the fields of 25 massive galaxy clusters, with a median of 658 objects per cluster field brighter than F814W = 25 mag (AB). A median of 463 of these galaxies per cluster are well detected (above $3 \sigma$ in at least 8 filters; 327 in 12, and 207 in 14, spanning the UV to the near-IR. Out photometry is given in the Appendix. 
\item Thanks to the optimized background measurement, source detection, and photometry techniques described in this paper, we obtain an approximately $30\%$ increase in the accuracy of photometric redshifts when compared with the previous CLASH photometry, with median absolute offsets between spectroscopic and photometric redshifts of $|(z_{\rm p} - z_{\rm s})| / (1 + z_{\rm s} ) = 0.030$. This photometric improvement is in addition to obtaining a much purer sample of cluster galaxies.
\item We are able to detect galaxies to $M^* + 4.5$ for all clusters and to $M^* + 7.5$ for the nearest clusters. The depth to which we can measure the luminosity function, combined with the large number of available filters with observations for each cluster, presents us with an unprecedented look at the growth of clusters across redshift. We see a passively evolving value of $M^*$ with redshift and no significant evolution in the population of faint galaxies (as measured by the slope of the luminosity function). 
\end{enumerate}

TC acknowledges support from a fellowship from the Michigan State Unversity College of Natural Science. T.C. and M.D. were supported by NASA/STScI grant {\it HST}-GO-12065.07-A. K.U. acknowledges support from the Ministry of Science and Technology of Taiwan through the grant MOST 106-2628-M-001-003-MY3. Results are partially based on ESO LP186.A-0798. This research has made use of the NASA/IPAC Extragalactic Database (NED) which is operated by the Jet Propulsion Laboratory, California Institute of Technology, under contract with the National Aeronautics and Space Administration. We used the cosmological calculator presented in \cite{2006PASP..118.1711W} in this work. Our work with Source Extractor was greatly aided by the guide from \citet{2005astro.ph.12139H}. T.C. thanks Thomas Hettinger and Ravi Jagasia for discussions on software implementation.  All of the data presented in this paper were obtained from the Mikulski Archive for Space Telescopes (MAST). STScI is operated by the Association of Universities for Research in Astronomy, Inc., under NASA contract NAS5-26555. Support for MAST for non-{\it HST} data is provided by the NASA Office of Space Science via grant NNX09AF08G and by other grants and contracts. 

\textit{Facilities}: {\it Hubble Space Telescope}/ACS, WFC3

\appendix{}

Our catalogs of observed photometry for all detected objects and rest-frame photometry for cluster members are provided online. We summarize the available photometry and photometric and spectroscopic redshift data for all galaxies in Table \ref{tab:clash_photometry}. For those galaxies we classified as members in Section \ref{sect:Members}, we present rest-frame photometry derived using iSEDfit in Table \ref{tab:clash_restframe_photometry}.
\begin{deluxetable*}{cll}
\tabletypesize{\scriptsize}
\tablecaption{Galaxy Properties}
\tablewidth{0pt}
\tablehead{ \colhead{Column Number} & \colhead{Column Name} & \colhead{Column Description}}
\startdata
 1 & Object ID & \\
 2 & Cluster & Cluster Name\\  
 3 & $\alpha_{2000}$ & Right Ascension (J2000) \\
 4 & $\delta_{2000}$ & Declination (J2000) \\
 5 & $X$ & $X$ pixel coordinate \\
 6 & $Y$ & $Y$ pixel coordinate \\
 7 & $a$ & Semimajor axis length (pixels) \\
 8 & $b$ & Semiminor axis length (pixels) \\
 9 & PA & Position angle (degrees) \\
10 & F225W Mag & Magnitude in F225W (mag) \\
11 & F225W Mag Error & Error in F225W magnitude (mag) \\
12 & F275W Mag & Magnitude in F275W (mag) \\ 
13 & F275W Mag Error & Error in F275W magnitude (mag) \\
14 & F336W Mag & Magnitude in F336W (mag) \\
15 & F336W Mag Error & Error in F390W magnitude (mag) \\
16 & F390W Mag & Magnitude in F390W (mag) \\
17 & F390W Mag Error & Error in F390W magnitude (mag) \\
18 & F435W Mag & Magnitude in F435W (mag) \\
19 & F435W Mag Error & Error in F435W magnitude (mag) \\
20 & F475W Mag & Magnitude in F475W (mag) \\
21 & F475W Mag Error & Error in F475W magnitude (mag) \\
22 & F555W Mag & Magnitude in F555W (mag) \\
23 & F555W Mag Error & Error in F555W magnitude (mag) \\
24 & F606W Mag & Magnitude in F606W (mag) \\
25 & F606W Mag Error & Error in F606W magnitude (mag) \\
26 & F625W Mag & Magnitude in F625W (mag) \\
27 & F625W Mag Error & Error in F625W magnitude (mag) \\
28 & F775W Mag & Magnitude in F775W (mag) \\
29 & F775W Mag Error & Error in F775W magnitude (mag) \\
30 & F814W Mag & Magnitude in F814W (mag) \\
31 & F814W Mag Error & Error in F814W magnitude (mag) \\
32 & F850LP Mag & Magnitude in F850LP (mag) \\
33 & F850LP Mag Error & Error in F850LP magnitude (mag) \\
34 & F105W Mag & Magnitude in F105W (mag) \\
35 & F105W Mag Error & Error in F105W magnitude (mag) \\
36 & F110W Mag & Magnitude in F110W (mag) \\
37 & F110W Mag Error & Error in F110W magnitude (mag) \\
38 & F125W Mag & Magnitude in F125W (mag) \\
39 & F125W Mag Error & Error in F125W magnitude (mag) \\
40 & F140W Mag & Magnitude in F140W (mag) \\
41 & F140W Mag Error & Error in F140W magnitude (mag) \\
42 & F160W Mag & Magnitude in F140W (mag) \\
43 & F160W Mag Error & Error in F160W magnitude (mag) \\
44 & ${\rm z}_{\rm b}$ & BPZ zb \\
45 & ${\rm z}_{\rm b}$ min & BPZ zbmin \\
46 & ${\rm z}_{\rm b}$ max & BPZ zbmax \\
47 & ${\rm z}_{\rm mk}$ & BPZ zml \\
48 & odds & BPZ Odds \\
49 & chisq & BPZ Chi-squared \\
50 & ${\rm z}_{\rm spec}$ & Spectroscopic redshift \\
51 & $\sigma {\rm z}_{\rm spec}$ & Spectroscopic redshift uncertainty\\
52 & ${\rm z}_{\rm spec}$ source & Source of spectroscopic redshift
\enddata
\tablecomments{This table is available in its entirety in machine-readable form.}
\label{tab:clash_photometry}
\end{deluxetable*}

\begin{deluxetable*}{cll}
\tabletypesize{\scriptsize}
\tablecaption{Rest-frame Photometry}
\tablewidth{0pt}
\tablehead{ \colhead{Column Number} & \colhead{Column Name} & \colhead{Column Description}}
\startdata
 1 & Object ID & \\
 2 & ${}^{0.0}u$ & ${}^{0.0}u$ magnitude \\
 3 & $\sigma ({}^{0.0}u)$  & ${}^{0.0}u$ magnitude uncertainty \\
 4 & ${}^{0.0}g$ & ${}^{0.0}g$ magnitude \\
 5 & $\sigma ({}^{0.0}g)$  & ${}^{0.0}g$ magnitude uncertainty \\
 6 & ${}^{0.0}r$ & ${}^{0.0}r$ magnitude \\
 7 & $\sigma ({}^{0.0}r)$  & ${}^{0.0}r$ magnitude uncertainty \\
 8 & ${}^{0.0}i$ & ${}^{0.0}i$ magnitude \\
 9 & $\sigma ({}^{0.0}i)$  & ${}^{0.0}i$ magnitude uncertainty \\
10 & ${}^{0.0}z$ & ${}^{0.0}z$ magnitude \\
11 & $\sigma ({}^{0.0}z)$  & ${}^{0.0}z$ magnitude uncertainty \\
12 & ${}^{0.0}$U & ${}^{0.0}$U magnitude \\
13 & $\sigma ({}^{0.0}$U)  & ${}^{0.0}$U magnitude uncertainty \\
14 & ${}^{0.0}$B & ${}^{0.0}$B magnitude \\
15 & $\sigma ({}^{0.0}$B)  & ${}^{0.0}$B magnitude uncertainty \\
16 & ${}^{0.0}$V & ${}^{0.0}$V magnitude \\
17 & $\sigma ({}^{0.0}$V)  & ${}^{0.0}$V magnitude uncertainty \\
18 & ${}^{0.0}$R & ${}^{0.0}$R magnitude \\
19 & $\sigma ({}^{0.0}$R)  & ${}^{0.0}$R magnitude uncertainty \\
20 & ${}^{0.0}$I & ${}^{0.0}$I magnitude \\
21 & $\sigma ({}^{0.0}$I)  & ${}^{0.0}$I magnitude uncertainty \\
22 & ${}^{0.0}$J & ${}^{0.0}$J magnitude \\
23 & $\sigma ({}^{0.0}$J)  & ${}^{0.0}$J magnitude uncertainty \\
24 & ${}^{0.0}$H & ${}^{0.0}$H magnitude \\
25 & $\sigma ({}^{0.0}$H)  & ${}^{0.0}$H magnitude uncertainty \\
26 & ${}^{0.0}$K & ${}^{0.0}$K magnitude \\
27 & $\sigma ({}^{0.0}$K)  & ${}^{0.0}$K magnitude uncertainty 
\enddata
\tablecomments{This table is available in its entirety in machine-readable form.}
\label{tab:clash_restframe_photometry}
\end{deluxetable*}

\bibliography{bibliography}
\end{document}